\documentclass[12pt]{article}
\usepackage{graphicx}
\usepackage{epstopdf}
\usepackage{amsmath}
\usepackage{amsfonts}
\usepackage{amssymb}
\usepackage{color}
\usepackage{mathrsfs}

\newcommand{\be}{\begin{equation}}
\newcommand{\ee}{\end{equation}}
\newcommand{\barr}{\begin{array}}
\newcommand{\earr}{\end{array}}

\usepackage{latexsym}
\usepackage{amsthm}
\usepackage{amsmath}
\usepackage{graphicx}
\usepackage{amssymb}
\usepackage{bbm}
\newcommand{\gsim}{\lower.7ex\hbox{$\;\stackrel{\textstyle>}{\sim}\;$}}
\newcommand{\lsim}{\lower.7ex\hbox{$\;\stackrel{\textstyle<}{\sim}\;$}}

\def\mpl{M_{\rm Pl}}
\newcommand{\bea}{\begin{eqnarray}}
\newcommand{\eea}{\end{eqnarray}}
\newcommand{\tdot}[1]{\dot{\ddot #1 }}
\newcommand{\comment}[1]{}
\newcommand{\Expect}[1]{\left\langle #1 \right\rangle}
\newcommand{\mbf}[1]{\mathbf #1}

\def\ep{\epsilon}
\def\SS{\mathcal{S}}
\def\T{\mathcal{T}}

\newcommand{\ba}{\begin{eqnarray}}
\newcommand{\ea}{\end{eqnarray}}
\newcommand{\nn}{\nonumber}

\def\k{{\bf k}}
\def\fnl{f_{NL}}
\def\hfnl{\hat f_{NL}}

\def\d{{\partial}}

\def\D{{\mathcal D}}
\def\O{{\mathcal O}}

\begin{document}


\setcounter{page}{1} \baselineskip=15.5pt \thispagestyle{empty}

\begin{flushright}
\end{flushright}
\vfil

\begin{center}

{\Large \bf New natural shapes of non-Gaussianity\\[0.1cm] from high-derivative interactions\\[0.3cm] and their optimal limits from WMAP 9-year data}
\\[0.7cm]
{\large Siavosh R.~Behbahani$^{a}$,   Mehrdad Mirbabayi$^{b}$,\\[0.2cm] Leonardo Senatore$^{c,d}$ and Kendrick M. Smith$^{e}$}
\\[0.7cm]

{\normalsize { \sl $^{a}$ Physics Department, Boston University, Boston, MA 02215}}\\
\vspace{.2cm}

{\normalsize { \sl $^{b}$  School of Natural Sciences, Institute for Advanced Study, Princeton, NJ 08540}}\\
\vspace{.2cm}

{\normalsize { \sl $^{c}$ Stanford Institute for Theoretical Physics, Stanford University, Stanford, CA 94306}}\\
\vspace{.2cm}

{\normalsize { \sl $^{d}$ Kavli Institute for Particle Astrophysics and Cosmology,\\ SLAC and Stanford University, Menlo Park, CA 94025}}\\
\vspace{.2cm}

{\normalsize { \sl $^{e}$ Perimeter Institute for Theoretical Physics, Waterloo, ON N2L 2Y5, Canada}}\\
\vspace{.3cm}

\end{center}

\vspace{.8cm}

\hrule \vspace{0.3cm}
{\small  \noindent \textbf{Abstract} \\[0.3cm]
\noindent Given the fantastic experimental effort, it is important to thoroughly explore the signature space of inflationary models. The fact that higher derivative operators do not renormalize lower derivative ones allows us to find a large class of technically natural single-clock inflationary models where, in the context of the Effective Field Theory of Inflation, the leading interactions have many derivatives. We systematically explore the 3-point function induced by these models and their overlap with the standard equilateral and orthogonal templates. We find that in order to satisfactorily cover the signature space of these models, two new additional templates need to be included. We then  perform the optimal analysis of the WMAP 9-year data for the resulting four templates,
finding that the overall significance of a non-zero signal is between 2--2.5$\sigma$,
depending on the choice of parameter space,
partially driven by the preference for nonzero $f_{NL}^{\rm orth}$ in WMAP9.
}
 \vspace{0.3cm}
\hrule

\begin{flushleft}
\end{flushleft}

\section{Introduction and Summary}

Inflation represents the leading candidate mechanism for the sourcing of the primordial fluctuations. Notwithstanding the great observational advances of the last two decades, with spectacular experiments that recently culminated in the Planck satellite or BICEP telescope, little is known about this primordial epoch. In an essential way, inflation can be defined as a primordial phase of quasi de Sitter epoch where time-translations are spontaneously broken~\cite{Cheung:2007st}, but little is known beyond this very general definition. The measurement of the primordial power spectrum of the density perturbations as observed with great precision in the Cosmic Microwave Background (CMB) or in Large Scale Structure (LSS) is ultimately traceable to the quadratic Lagrangian of the fluctuations during the primordial quasi de Sitter phase. It would be much more interesting to test the interacting structure of the theory, in order to understand the nature  of the physics that drove inflation and its  connections with other fundamental interactions. The leading and most general way to test the interacting structure of the inflationary field is by measuring the non-Gaussian signal that is induced by its self-interactions. Not only do non-Gaussianities offer a signal that is directly associated to the interacting structure of the theory, but they represent such a rich signal from the observational point of view so that, if detected, a plethora of interesting measurements would be possible beyond the first detection. For example, if we were to detect a primordial 3-point function, we would be subsequently curious to know how strong is the signal as a function of the triangular configurations in momentum space, the so-called shape of the non-Gaussianities; or if there is also a non-trivial 4-point function.

So far, limits on inflationary 3-point function have been focussed on two different approaches. The first is based on providing templates for 3-point functions that are matched against the data; while the second attempts trying to reconstruct a generic 3-point function from the data. The advantage of the first method is that it can focus directly on theoretically motivated models over which one can perform an optimal analysis. It has however the disadvantage of potentially missing a signal present in the data simply because it was not looked for. So far, this first approach has been used to search for the equilateral~\cite{Creminelli:2005hu} and orthogonal~\cite{Senatore:2009gt} templates from single field inflation~\cite{Alishahiha:2004eh,Cheung:2007st}, as well as the local template~\cite{Komatsu:2003iq} from multi field inflation~\cite{Lyth:2002my,Zaldarriaga:2003my,Senatore:2010wk}. The second approach~\cite{Fergusson:2009nv} is instead based on reconstructing the primordial signal present in the data by matching it to a basis of functions that can cover, at least in principle, any potential signal. This second approach has the advantage of being very thorough, though it has the disadvantage of being rather suboptimal, as the significance of a signal can be diluted away as many independent shapes are matched to the data.

So far, the most constraining search for non-Gaussianities is provided by the analysis of the Planck team~\cite{Ade:2013ydc}, which finds no evidence of non-Guassianities. It should be stressed however that this limit is still rather weak: the skewness of the distribution of the primordial fluctuations is constrained to be smaller than about $10^{-3}$. Apart from the awe that is associated to us, mankind, being to be able to say anything about the first instants of the universe, such a limit is not very strong from the particle physics point of view: it constrains inflation to be more or less as interacting as the electron in Quantum ElectroDynamics, or as the Pion at energies of order of its mass. It would be clearly very interesting to be able to further constrain the level of non-Gaussianities by one or two orders of magnitude, a sensitivity that the recently developed Effective Field Theory of Large Scale Structures~\cite{Carrasco:2012cv} has shown the potential to achieve with LSS surveys in the next decade; but also to explore in an optimal way all possible signals potentially produced by inflation.

The purpose of this paper is to present a more detailed exploration of the 3-point functions that are theoretically predicted by the single field inflationary models. The study of the phenomenology of inflation is greatly simplified by the so-called Effective Field Theory of Inflation~(EFTofI)~\cite{Cheung:2007st,Senatore:2010wk}. By postulating that inflation is the theory associated to the spontaneous breaking of time-translation in a quasi de Sitter epoch, it allows a description of the dynamics of the fluctuations during inflation, the so-called $\pi$ field, without the need of describing the background solution, which is largely irrelevant for the purpose of predicting the observations. With the EFTofI, the analysis of the signals that are producible during inflation becomes much simpler.  We will work in the context of the EFTofI as applied to single field inflation~\footnote{The same approach has been extended to cover multiple fields~\cite{Senatore:2010wk} and dissipative effects~\cite{LopezNacir:2011kk} in inflation. The formalisms that lead to the Effective Field Theory of Inflation were first constructed in~\cite{Creminelli:2006xe} to study alternative cosmology and violation of the null energy condition, and in particular dark energy. The application to dark energy has been subsequently re-taken and further developed in~\cite{Gubitosi:2012hu}, which gave the name to this application of this formalism already started in~\cite{Creminelli:2006xe} as Effective Field Theory of Dark Energy. Notice that while the presence of an additional degree of freedom is a necessary requirement in Inflation, in the case of the current acceleration of the universe this fact needs to be postulated.}, where we also impose an approximate shift symmetry that forces each power of $\pi$ to appear with a derivative acting on it and that makes therefore any interaction a higher-dimensional one. At leading order in the derivative interactions,  there are two operators: $\dot\pi^3$ and $\dot\pi({\partial}_i\pi)^3$, and the signals that can be produced are well described by the equilateral~\cite{Creminelli:2005hu} and orthogonal templates~\cite{Senatore:2009gt}.  Our study kicks off from a simple field theoretic fact that we explain in section~\ref{sec:naturalness}: a theory where the leading interactions have a certain number of derivatives acting on each $\pi$ will not induce, under renormalization, interactions with a smaller number of derivatives acting on each $\pi$. This allows us to construct a very large number of technically natural inflationary models where the leading interactions have many derivatives acting on each~$\pi$, while the interactions with a smaller number of derivatives are consistently suppressed. We also discuss a symmetry that can be imposed on the theory and that partially justifies these findings.

When applied to the EFTofI, with its own additional constraints coming from the nonlinear realization of Lorentz invariance, this sort of non-renormalization theorem opens up a plethora of technically natural inflationary models, each one with its own leading higher derivative operators, and distinguished by its own non-Gaussian signals. However, as we notice in section~\ref{indi}, many of these operators lead to identical shapes of non-Gaussianities at tree level, a fact that can be made explicit by performing appropriate field redefinitions. The degeneracy of signals becomes even more striking after we realize that many models which include operators that produce in-principle-novel shapes of non-Gaussianities, in reality induce a signal that is practically indistinguishable from a linear combination of standard templates, as we discuss in section~\ref{sec:new-templates}.

As we move to higher and higher numbers of derivatives, the complexity of the models becomes larger and larger. We therefore develop some statistical ways of exploring the parameter space of models, and include a new template only if there exist a natural model for which an order one fraction of the parameter space produces a signal that is not covered by the standard templates (we take a fraction to be order one if it is larger than about 10\%). This exploration concludes, since, as we move to higher derivative operators and ask for them to produce a detectable signal, the mass scale suppressing them becomes lower and lower, becoming dangerously close to Hubble scale. We, reasonably but somewhat arbitrarily, decide to stop at the level of 9-derivative operators, and decide that to consistently discuss models beyond this number of derivatives new light degrees of freedom should be included.

The result of this analysis is that we find that up to 6-derivative operators, there is no need to add any new template beyond the standard equilateral and orthogonal ones: $F_{\rm equil}$ and $F_{\rm orth}$. At 7-derivatives, we need to add one template, that we call $F_{\rm 7 der}$. These three templates are sufficient at 8-derivative level, while at the level of 9 derivatives we need to add one more template, labelled $F_{\rm 9 der}$. In summary, we find that a total of four templates is sufficient to analyze the signal produced by all technically natural single clock inflationary models.

Finally, in the last section~\ref{sec:analysis}, we perform the optimal analysis of the WMAP 9 year data for all the four templates, two of which have never been matched to the data before. The ``bottom line'' results are given by
\ba
f_{NL}^{\rm eq} &=& 51 \pm 136 \hspace{1.94cm} \mbox{($-221 < f_{NL}^{\rm eq} < 323$ at 95\% CL)}  \nn \ ,\\
f_{NL}^{\rm orth} &=& -245 \pm 100 \hspace{1.5cm} \mbox{($-445 < f_{NL}^{\rm orth} < -45$ at 95\% CL)}  \nn \ ,\\
f_{NL}^{\rm 7der} &=& -34 \pm 56 \hspace{1.85cm} \mbox{($-146 < f_{NL}^{\rm 7der} < 78$ at 95\% CL)}  \nn \ , \\
f_{NL}^{\rm 9der} &=& 30 \pm 16 \hspace{2.12cm} \mbox{($-1 < f_{NL}^{\rm 9der} < 62$ at 95\% CL)}  \nn\ .
\ea
We also perform a joint analysis of multiple templates,
and find that deviations from Gaussianity are favored at 2--2.5$\sigma$
depending on the choice of parameter space.
This is not statistically significant, but could clearly become so with the Planck data. 
The significance we find in the WMAP 9-year data is partially driven by the preference for nonzero $f_{NL}^{\rm orth}$ in WMAP9,
which is known to decrease in Planck~\cite{Ade:2013ydc}.
We do not analyze Planck data in this paper, since Monte Carlo simulations of the foreground-cleaned Planck maps 
(used by the Planck team in the non-Gaussianity paper~\cite{Ade:2013ydc}) are not yet publicly available, 
and would be impractical to construct due to the complexity of the Planck noise model.
It will be interesting to perform an analysis of Planck data in the future. 

Higher derivative operators were already considered in the context of the EFTofI in~\cite{Bartolo:2010bj}, without giving explanations on how the models considered there would be technically natural, and in~\cite{Creminelli:2010qf}, where the authors use the Galileian symmetry to suppress the lower derivative operators. In both cases the authors stop at 6 derivative level, which we argue is already covered by the standard templates.

\section{The Effective Field Theory of Inflation\label{sec:naturalness}}

Let us begin with a review of the effective field theory of inflation (EFTofI)~\cite{Cheung:2007st}. The starting point is to choose constant time slices to coincide with constant clock slices (the so-called unitary gauge). Then, the most general single clock field theory of inflation can be described by the Einstein-Hilbert action plus terms that contain metric fluctuations and their derivatives, and are invariant under all but time diffeomorphisms \cite{Cheung:2007st}:
 \begin{eqnarray}
 S = \!\!\!\int  \! d^4 x \; \sqrt{- g} \Big[ \frac12 M_{\rm
Pl}^2 R + M_{\rm Pl}^2 \dot H
g^{00} - M_{\rm Pl}^2 (3 H^2 + \dot H)\nonumber\\
\quad \!\!\!+ \frac{1}{2!}M_2(t)^4(g^{00}+1)^2
+ \frac{1}{3!}M_3(t)^4 (g^{00}+1)^3+\cdots \Big]\ .
\label{S}
\end{eqnarray}
Here, the coefficients of first three terms, which are the only ones which start linear in the fluctuations, are determined in terms of cosmological history, $H$ and $\dot H$, but those of the higher order terms are expected to be generic functions of time. One can restore a non-linearly realized full reparametrization invariance by performing a time-diff. $t\to t+\pi$ in \eqref{S}, and promoting $\pi$ to a new field which non-linearly shifts under time-diffs. It is then easily seen that on a quasi-de Sitter background (where $\ep=-\dot H/H^2 \ll 1$) and at energies of order and higher than $H$, $\pi$ decouples from metric perturbations and captures the dynamics of the inflaton field~\cite{Cheung:2007st}. The relevant part of the action \eqref{S} then becomes
\begin{eqnarray}\label{S1}
&&S_{\rm \pi} = \\ \nonumber
&&\ \  \int d^4 x   \sqrt{- g} \left[ -M^2_{\rm Pl}\dot{H} \left(\dot\pi^2-\frac{ (\partial_i \pi)^2}{a^2}\right)
 + 2 M^4_2
\left(\dot\pi^2+\dot{\pi}^3-\dot\pi\frac{(\partial_i\pi)^2}{a^2}
\right) -\frac{4}{3} M^4_3 \dot{\pi}^3 +\ldots\right] . 
\eea
The interaction terms in the above action, which lead to non-Gaussian correlation functions, are constrained by the requirement of non-linear realization of time diffs. That is to say, the fact that the most general action for $\pi$ must arise from \eqref{S} enforces that (a) all terms are linearly rotationally invariant, (b) once terms are non Lorentz invariant at linear level, there are connections between terms of different order in $\pi$ so that Lorentz invariance is non-linearly realized (e.g. the terms multiplied by $M_2^4$ in \eqref{S1}), and (c) interactions that contain $\pi$ without derivative acting on it (e.g. $\pi\dot\pi^2$) must be multiplied by time derivatives of the time-dependent coefficients in the Lagrangian. However, it is technically natural to imagine that these derivatives are  proportional to slow-roll parameters and small on a quasi-de Sitter background. It therefore suffices to concentrate on the derivative interactions~\footnote{Resonant non-Gaussianity~\cite{Flauger:2010ja} models may seem as an exception to this rule, however they are shown to have non-Gaussian signals that are subleading with respect to the signal in the power spectrum~\cite{resonant}~(but see \cite{Daniel_Siavosh}).}. 

As it is shown explicitly in \eqref{S1}, the cubic interactions that satisfy the above constraints and have the minimum number of derivatives, i.e. three, are $\dot\pi^3$ and $\dot\pi \pi_{,i}^2$. However, it is crucial to be able to fully explore the space of non-Gaussian signatures of inflation,  as inflation is one of the pillars of the physics beyond the standard model, and non-Gaussianities are the probe of the interaction structure of the theory. It is also important to do this in light of the huge experimental effort in understanding the nature of the primordial fluctuations, both in the CMB and in Large Scale Structures. For this reason, in this paper we realize that higher derivative interactions can be the dominant source of non-Gaussianity, and decide to focus on them.  Before even beginning to study these models, one should ask how, without tuning, it is possible for the higher derivative operators to be the leading ones, and not having observationally-more-important lower derivative operators. The first answer to this question is to actually realize that a theory where the leading interaction operator is a higher derivative one is technically natural. We explain this simple fact here for Lorentz invariant theories, and we discuss in App.~\ref{natural} how the argument extends to the EFTofI. Let us consider the following Lorentz Invariant theory of a scalar field in Minkowski space with a single higher derivative quartic interaction:
\be\label{eq:theory1}
S=\int d^4x \left[\frac{1}{2}{\partial}_\mu\phi{\partial}^\mu\phi+ \frac{1}{\Lambda^8} (\d_\nu\d_\mu\phi\d^\nu\d^\mu\phi)^2\right]\ .
\ee
At loop level, even in the case in which we integrate the internal momenta all the way up to the unitarity bound $\Lambda$, it is quite straightforward to realize that  only operators with an higher number of derivatives acting on each $\phi$ will be generated, such as $ \frac{1}{\Lambda^{10}} (\d_\rho\d_\nu\d_\mu\phi\d^\rho\d^\nu\d^\mu\phi)({\partial}_\nu\d_\mu\phi\d^\nu\d^\mu\phi)$, but {\it no} operator with a lower number  of derivatives acting of $\phi$, such as $ \frac{1}{\Lambda^4} (\d_\nu\phi\d^\nu\phi)^2$. The proof of this fact at all orders in perturbation theory is rather straightforward. Let us imagine to evaluate the connected correlation function of $\langle \phi(k_1)\ldots\phi(k_n)\rangle_c$ by inserting an arbitrary number of times the interaction vertex. Each of the external $\phi(k_i)$'s will be contracted with one of the $\phi$'s in the vertex, which, being acted upon by two derivatives, will give rise to a factor of $k_{i}^\mu k_i^\nu$ (see Figure~\ref{fig:technical-diagram}). Neglecting the indexes, this means that the regularized (that is before renormalization) expression of the 1PI correlation function is 
\be
\langle \phi(k_1)\ldots\phi(k_n)\rangle_{\rm 1PI}= k_1^2 \ldots k^2_n\; f\left(k_1,\ldots k_n; \Lambda,\Lambda_{\rm cutoff}\right)
\ee
where $\Lambda_{\rm cutoff}$ is the parameter of some form of UV regulator.  If at some order $n$ in perturbation theory all divergent sub-diagrams that appear in the calculation of this correlation function are systematically renormalized, the divergent part of $f$ (in the limit $\Lambda_{\rm cutoff}\to \infty$) is guaranteed to be local. This means that, as we send {\it any} of the external momenta $k_i$ to zero, the correlation function vanishes at least as $k_i^2$. Therefore {\it any} counterterm that will be needed to renormalize the theory will have at least two derivative acting on each $\phi$, as we wanted to show. In formulas, this means that a technically natural version of the theory in (\ref{eq:theory1}) is
\bea\label{eq:theory2}
&&S=\int d^4x \left[\frac{1}{2}\d_\mu\phi\d^\mu\phi+\frac{c_2}{\Lambda^2} \d_\nu\d_\mu\phi\d^\nu\d^\mu\phi+c_3 (\d_\mu\d^\mu\phi)^2\right.\\ \nonumber
&&\quad\qquad\qquad\left.  +\frac{1}{\Lambda^8} (\d_\nu\d_\mu\phi\d^\nu\d^\mu\phi)^2+\frac{c_4}{\Lambda^8}(\d_\mu\d^\mu\phi)^4\right.\\ \nonumber
&&\left.\quad\qquad\qquad+ \frac{c_5}{\Lambda^{10}} (\d_\rho\d_\nu\d_\mu\phi\d^\rho\d^\nu\d^\mu\phi)(\d_\nu\d_\mu\phi\d^\nu\d^\mu\phi)+\ldots\right]\ ,
\eea
where $\ldots$ represents higher dimension operators and $c_i$ are order one numerical coefficients~\footnote{The field theoretic realization above makes it interesting to explore the possibility of models that have an higher unitary bound than in~(\ref{eq:theory2}) and that, in the low energy regime, reproduce our technically natural models. In fact, it is easy to realize that if we let the scalar field $\phi$ interact with a massive high spin particles, rotational invariance will force the leading interactions to have many derivatives acting at least on one of the $\phi$'s, so that, once we integrate out the massive particle, we can imagine to be left with a model like in~(\ref{eq:theory2}). The problem with this candidate UV complete model is that, due to the well known difficulties of UV completing theories with massive high spin particles,  the unitarity bound of the new model turns out to be not higher than the one in~(\ref{eq:theory2}), at least for the cases we explored. A well known theory which is UV complete and contains high spin particles is string theory, and it is therefore tempting to try to UV complete these models directly into a string theory. We leave this study to future work; we, who are most interested in the connection with data, content ourselves in this paper with the Effective Field Theory.}. The purpose of this paper is to apply this novel quantum field theory fact to the theory of inflation, by studying the theoretical and observational implications of higher derivative operators. Being inflation a theory where Lorentz invariance is non-linearly realized, there are some constraints in applying this quantum field theory fact, as discussed in detail in Appendix~\ref{natural}.

The fact that at least a fraction of the lower derivative operators are not generated under renormalization by the higher derivative ones can be justified by imposing some symmetries that offer a generalization of the shift symmetry of the Galileon~\cite{Nicolis:2008in}. Indeed, one can impose the theory to be invariant under a shift
\be
\phi(x)\quad\to\quad \phi(x) + C_{\mu_1\ldots\mu_n}\; x^{\mu_1}\ldots x^{\mu_n}\ .
\ee
If the tensor $C$ is `traceless', in the sense that $C^\mu{}_{\mu\mu_2\mu_3\dots}=0$, then the kinetic term is left invariant, and the leading derivative interactions that are obviously allowed are those with $n+1$ derivatives acting on each $\phi$. The case $n=1$ is the so-called Galileon, for which this symmetry does not forbid some lower derivative operators, such as $(\d\phi)^2(\d_\mu\d^\mu\phi)$, an operator that is connected to potential superluminal propagation of the fluctuation~\cite{Adams:2006sv}. Similar operators that involve $\phi$-legs with only $n$ derivatives acting on them may exist also for $n>1$~\footnote{Since this symmetry argument is alternative to our perturbation theory argument, we will not discuss it for the non-relativistic case, though it is possible that a non-relativistic generalization might be found.}.

\begin{figure}[t]
\begin{center}
\includegraphics[width=10 cm, height= 8 cm]{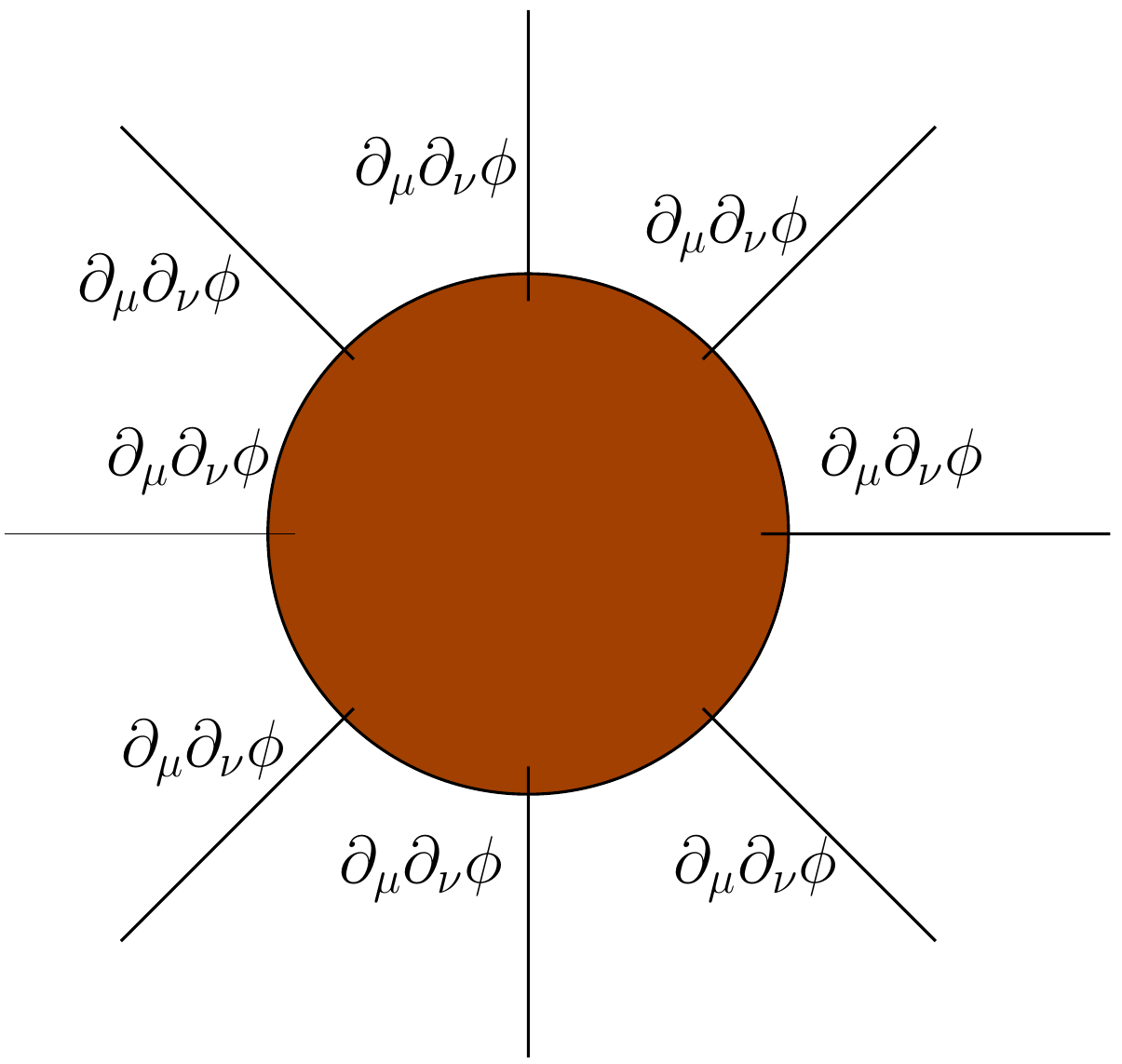}
\end{center}
\caption[]{The non-renormalization theorem that we explain in the text: it is impossible to generate operators with a lower number of derivatives acting on each $\phi$-leg.}
\label{fig:technical-diagram}
\end{figure}

Now that we have established that we can consider higher derivative operators as the leading ones, there is a next natural question related to the fact that it seems that now we have to analyze the observational consequences of an infinite number of operators. There is a very natural limit beyond which it is not worth to proceed. Consider a general cubic interaction with $n$ derivatives, schematically,
\bea
\label{L3}
{\cal L}_3 = \frac{1}{\Lambda^{n-1}}\partial^n\pi_c^3\,,
\eea
where $\pi_c$ is the canonically normalized field and $\partial$ denotes either time or space derivative. $\Lambda$ is some mass scale representing the unitarity bound of the theory, and the high energy cutoff of the theory $\Lambda_{UV}$ must satisfy $\Lambda_{UV}\leq \Lambda$~\footnote{\label{cs} When $c_s<1$ some care should be taken in determination of the suppression scale $\Lambda$. Namely, one should first replace $\vec{x}\to \vec{\tilde{x}}= c_s\vec{x}$ in the EFTofI action of goldstone mode $\pi$ and then canonically normalize $\pi_c = (-2\mpl^2\dot H c_s)^{1/2}\pi$ to obtain the suppression $\Lambda$ in \eqref{L3} (see \cite{4point}).}. The non-Gaussianity generated by \eqref{L3} can be estimated by comparing ${\cal L}_3$ with the kinetic term ${\cal L}_2 = \dot\pi_c^2$ at the energy scale $H$, which yields
\bea 
\label{fnl}
f_{NL}\zeta \sim \frac{{\cal L}_3}{{\cal L}_2} \sim \left(\frac{H}{\Lambda}\right)^{n-1}\,,
\eea
where $\zeta = H^2/\sqrt{-4\mpl^2\dot Hc_s}\simeq 5\times 10^{-5}$ is the amplitude of scalar curvature fluctuations. It follows from \eqref{fnl} that in order to get a fixed level of non-Gaussianity ($f_{NL}=\rm{const}$), when increasing the number of derivatives, $\Lambda$ and consequently $\Lambda_{UV}$ must approach $H$. On the other hand, in order to have a reasonably local and weakly coupled theory at Hubble scale we must stop at some point. We, somewhat arbitrarily, choose  $\Lambda \geq 3H$ which for $f_{NL}\zeta\sim  10^{-4}$ leads to $n\leq 9$. This is the maximum number of derivatives we wish to consider, even though, at least in principle, the analysis can be continued further. To consistently go beyond, however, we believe one should add an additional light degree of freedom.

It is clear from \eqref{fnl} that in the presence of a lower derivative operator suppressed by the same mass scale $\Lambda$ the higher derivative operators are irrelevant and negligible. Therefore, the interaction \eqref{L3} is important only if lower derivative operators are suppressed by a larger mass scale:
\bea
\label{hierarchy}
{\cal L'}_3 = \frac{1}{(\Lambda')^{n-m-1}}\partial^{n-m}\pi_c^3\,,
\eea
with ${\Lambda'}^{n-m-1} > \Lambda^{n-1}/H^m $. We will explain in Appendix \ref{natural} when and how such a hierarchy of scales can be consistently realized in EFTofI, resulting in technically natural models in which high-derivative operators lead to large non-Gaussianity. 

In the next section, starting from 4-derivative interactions, we increase the total number of derivatives and check for shapes of 3-point function that are significantly different from the ones analyzed so far. We will encounter some explicit consequences of the requirement of technical naturalness in our investigation. 

\section{\label{indi} Independent higher derivative operators}

When studying the non-Gaussian signature induced by some higher derivative operators, it is important to realize that there are cases in which the induced shape is equal to a linear combination of shapes generated by lower derivative ones. For example, consider the following four-derivative interactions
\bea
\label{4der}
\ddot\pi\dot\pi\dot\pi\ ,\qquad \d^2\pi\pi_{,i}\pi_{,i}\ ,\qquad \d^2\pi \dot\pi^2\,.
\eea
Since we are dealing with interactions with several derivatives, we adopt the shorthand $\pi_{,i}=\d_i\pi$ for spatial derivatives. According to the discussion of Appendix \ref{natural}, there are natural models in which these are the leading operators and 3-derivative interactions are either absent or suppressed by $H/\Lambda$. For instance, one can have a model in which the cubic Lagrangian is given by
\bea
\label{4derL3}
\mathcal{L}^{(3)}= \frac{1}{\Lambda^3}(a_1 \d^2\pi_c\dot\pi_c^2+ a_2 H \dot\pi^3)\ ,
\eea
with $a_i$'s being order one numerical coefficients, and where the suppression of the second term compared to the first is natural since they have different $t\to -t$ symmetry, which is softly broken by the background cosmology. 

However, it is easy to see that all interactions in \eqref{4der} reduce to linear combinations of $H\dot\pi^3$ and $H\dot\pi\pi_{,i}^2$ via integration by parts, and ignoring terms that are proportional to the variation of the quadratic Lagrangian: $\delta\mathcal{L}^{(2)}/\delta\pi \propto \ddot\pi+3H\dot\pi-c_s^2\d^2\pi$. The latter means that a given interaction term can be removed by performing a field redefinition that indeed removes this term from the cubic action and introduces new quartic and higher order interactions. In calculating the correlation functions of $\pi$ (or of the curvature perturbation $\zeta$, which is equal to $\zeta=-H\pi$ at linear level), one has to keep track of these field redefinitions~\cite{Maldacena:2002vr}, but since for us they always contain derivatives of fields, their contribution to the late time correlators exponentially decays, and we can therefore neglect it~\footnote{For instance, in \eqref{4derL3}, $\dot\pi_c^2\d_i^2\pi_c$ can be replaced by $\dot\pi_c^2(\ddot\pi_c+3H\dot\pi_c)$ via the redefinition $\pi_c \to \pi_c+\dot\pi_c^2/\Lambda^3$, and $\dot\pi_c^2\ddot\pi_c$ is a total derivative which can be replaced by $-3H\dot\pi_c^3$. The quadratic term in the field redefinition $\dot\pi_c^2$ gives an exponentially small contribution for the late time correlation functions. This means that the operator $\dot\pi_c^2\d_i^2\pi_c$ induces the same shape of the 3-point function as the operator $\dot\pi^3$.}.

One can use similar manipulations and induction at higher derivative level to show that the shapes of all interactions that respect the shift symmetry $\pi\to \pi+c$ can be reduced to linear combinations of the shapes induced by the following terms
\bea
\label{t}
&\partial_t^n\pi\partial_t^n\pi\partial_t^m\pi\,,&\quad\text{with}\quad n\geq m\geq 1\,,\\
\label{i}
&\partial_t^{2n-1}\pi \pi_{,i}\pi_{,i}\,,&\quad\text{with} \quad n\geq 1.
\eea
We show this in Appendix~\ref{app:reduction}.
Hence, the shapes of these operators form a complete basis for 3-point function in single-field inflation as long as the kinetic term is of the form $\dot\pi^2-c_s^2\pi_{,i}^2$. 

Let us now proceed to operators with at most five derivatives. According to (\ref{t},\ref{i}) there are two new independent operators at this level; namely, $\ddot\pi^2\dot\pi$ and $\tdot\pi\pi_{,i}^2$. However, theoretical considerations forbid both of them from being the leading interaction since if 
\bea
\mathcal{L}^{(3)}\supset \ddot\pi_c^2\dot\pi_c/\Lambda^4,
\eea
then the loops of this operator generate $\dot\pi_c^3/\Lambda^2$, which, without fine-tuning, dominates the 3-point function. On the other hand $\tdot\pi\pi_{,i}^2$ can only arise from quadratic unitary gauge operators such as
\bea
\delta g^{00} {\partial^0}\d^0g^{00} ,
\eea
which inevitably induce the high derivative term $\tdot\pi\dot\pi$ in the kinetic term. It is easy to see that in order for the non-Gaussianity to be detectably large the high time-derivative correction to the kinetic term must be dominant. This introduces an additional light degree of freedom, which is a ghost. So, we do not have new shapes at 5-derivative level either~ \footnote{In fact, even other 5-derivative interactions that are reducible and not included in \eqref{t} and \eqref{i} do not pass our theoretical criteria, because in all of them at least one $\pi$ must have only one derivative. If this is a $\dot\pi$, then loops generate $\dot\pi^3$, and if it is $\pi_{,i}$, the operator can only come from a quadratic unitary gauge operator.}.

Once the 6-derivative operators are considered, we not only get the new shape $\ddot\pi^3$ from~\eqref{t}, but also the 5-derivative terms can now arise in a technically natural way. For instance, one can have a model
\bea
\label{6derL3}
\mathcal{L}^{(3)}= \frac{1}{\Lambda^5}(a_1 \ddot\pi_c^3+ a_2 H \ddot\pi_c^2\dot\pi_c+a_3 H^3 \dot\pi_c^3)\ ,
\eea
with $a_i$'s of order one, where now because of the $H/\Lambda$ suppression of $\ddot\pi^2\dot\pi$, its loops generate $\dot\pi^3$ with $H^3/\Lambda^3$ suppression, which leads to comparable level of non-Gaussianity as the higher derivative term. Moreover, there are 6-derivative interactions that come from cubic unitary gauge operators, and upon reduction to (\ref{t},\ref{i}) have a non-zero coefficient of $\tdot\pi\pi_{,i}^2$ (e.g. $\ddot\pi\pi_{,ij}^2$ which comes from $\partial^0g^{00}\delta E_{ij}^2$). One, therefore, expects to have three new shapes at this level. Similar arguments can be used to show that there are no new shapes at 7-, two at 8-, and two at 9-derivative levels. 

While in a strict mathematical sense this is the right conclusion we will next argue that in practice the number of new shapes is much smaller since they are often very similar and observationally indistinguishable.

\section{New Templates\label{sec:new-templates}}

Consider a given cubic interaction $I$ (for instance $I$ can be $\dot\pi^3$). The shape of 3-point function, $s_I(k_1,k_2,k_3)$, produced by $I$ is defined in terms of the late time momentum space correlator
\bea
\Expect{\pi_{\mbf{k}_1}\pi_{\mbf{k}_2}\pi_{\mbf{k}_3}}= s_I(k_1,k_2,k_3)(2\pi)^3\delta^3(\mbf{k_1}+\mbf{k_2}+\mbf{k_3})\ ,
\eea 
which because of invariance of the theory under spatial translations and rotations is only a function of the lengths of the momenta of the three modes. The approximate scale invariance of the quasi-de Sitter background implies that $s_I(k_1,k_2,k_3)\simeq s_I(1,k_2/k_1,k_3/k_1)/k_1^6$; moreover, it is obviously invariant under permutations of $\{k_1,k_2,k_3\}$. Correlation functions of $\pi$ can easily be transformed into those of the gauge invariant scalar curvature $\zeta$ using $\zeta =-H\pi$ to leading order in the slow roll parameters.

To constrain an inflationary model that contains $I$ one can use $s_I$ as a template and convolve it with the data. However, two different shapes can be correlated, in which case a constraint on one also constrains the other. To quantify this concept one defines the inner product of two shapes $s$ and $s'$ according to~\cite{Babich:2004gb}
\bea
\label{dot}
(s,s')=\int_{1/2}^1 \!\! dx_2\int_{1-x_2}^{x_2}\!\!\!\!\!dx_3\;x_2^4x_3^4\; s(1,x_2,x_3)\;s'(1,x_2,x_3)\ ,
\eea
and the norm of a shape as $|s|=\sqrt{(s,s)}$. The cosine between two shapes, which is a normalized measure of their correlation, is then
\bea
\label{cosine}
\cos(s,s')=(s,s')/(|s||s'|)\,.
\eea

This notion can readily be generalized to the correlation (or maximum cosine) of a shape $s$ with a set of shapes (or templates) $\T=\{t_i\}$, which is defined as $\max\{\cos(s,\sum \alpha_i t_i)$ $|\forall\{\alpha_i\}\}$; it is given by 
\bea
\label{theta}
\cos \theta_{s,\T}\equiv\cos(s,\T)= \sqrt{\mbf{v}^T \mbf{A}^{-1}\mbf{v}}/|s|\ ,
\eea
where $v_i=(s,t_i)$ and $A_{ij}=(t_i,t_j)$. The parallel $s_\parallel$ and perpendicular $s_\perp$ components of $s$ with respect to $\T$ are defined in terms of $\theta_{s,\T}$ in an obvious way.

Let us check the correlation of the three new shapes of $\ddot\pi^3,\tdot\pi\dot\pi^2,$ and $\tdot\pi\pi_{,i}^2$, which were found at 6-derivative level, with the shapes of the lowest order interactions $\dot\pi^3$ and $\dot\pi\pi_{,i}^2$. Rather surprisingly, we find that all the three shapes are more than 0.99 correlated with the set $\{s_{\dot\pi^3},s_{\dot\pi\pi_{,i}^2}\}$. This implies that unless we consider linear combinations of the operators with abnormally large coefficients, the new models are practically indistinguishable from the 3-derivative ones. 

However, it is too soon to draw this conclusion since the process of reduction to the form (\ref{t},\ref{i}) does indeed produce large numerical factors, specially at high derivative levels. On the other hand, it seems unnatural to allow linear combinations of operators with arbitrarily large coefficients. One is, therefore, faced with the question of which linear combinations are realistic and which ones are not. To answer this question, we make from here on the explicit assumption that the coefficients of the operators in unitary gauge, apart from appropriate mass parameters, vary in an order one range.  Thus, we avoid any reduction to (\ref{t},\ref{i}) but at each derivative level we list all cubic interactions that can be a dominant source of non-Gaussianity in a technically natural model. Call the set of all these shapes $\mathcal{U} =\{s_i\}$. Each technically natural model has a 3-point function which is a linear combination of a subset $\SS\subset \mathcal{U}$, and our assumption about the coefficients of the unitary-gauge operators implies that the coefficients in these linear combinations are also~$\mathcal{O}(1)$.  

We then enlarge the set of linearly independent templates $\T$ until {\it almost all} of the technically natural inflationary models with order one coefficients are {\it covered}. Note that making subjective choices to define `almost all' and `being covered' is unavoidable unless we take $\T$ to contain all operators that are given by (\ref{t},\ref{i}). Our criterion for a shape~$s$ to be considered covered is $s_\parallel>s_\perp$ or equivalently $\cos \theta_{s,\T}>0.7$. To have a notion of `almost all', one first needs to put a measure on the parameter space of the theory, for instance, by postulating that all order one coefficients are equally likely, and then set a threshold that we take to be of order 10\%. That is, if more than 90\% of the parameter space has a correlation larger than $0.7$ with $\T$, we do not enlarge $\T$.

To see an explicit example of how this program can in principle be accomplished, let us review the construction of the orthogonal shape \cite{Senatore:2009gt}. Consider the lowest derivative inflationary models with cubic Lagrangian $\alpha_1\dot\pi\pi_{,i}^2+\alpha_2 \dot\pi^3$, and suppose $\T$ contains only the equilateral shape, which is almost identical to the shape of $\dot \pi \pi_{,i}^2$. Since $\dot\pi^3$ is not exactly aligned with the equilateral shape (they have a cosine of $0.95$), a fraction of about 10\% of the parameter space is not covered by  $\T=\{{\rm Equil.}\}$. Very explicitly: as one varies the coefficients $\{\alpha_1,\alpha_2\}$ in an order one range, in a fraction of about 10\% the overall shape of the model has cosine of less than $0.7$ with the equilateral shape. Thus, it is reasonable to enlarge $\T$ by adding an orthogonal template based on $s_{\dot\pi^3}$~\footnote{\label{(-1,1)}In practice we first normalize shapes and then vary the coefficients in the range $(-1,1)$. This differs from varying the coefficients of unitary gauge operators in $(-1,1)$ by the ratio of the norm of different shapes, which is expected to be usually of order one. The advantage is that all shapes are treated on equal footing, and the application of the geometric analysis of Appendix \ref{geo} is much easier.}.

While the procedure to determine the number of shapes to analyze that  we just outlined is satisfactory (at least to us), as we move on to consider higher derivative operators, the number of operators and as a consequence the number of technically natural models rapidly grows, and the above direct approach becomes less and less feasible (recall that we avoid reduction to (\ref{t},\ref{i}) and consider almost all interactions). We, therefore, use an alternative method that allows us to carry out the analysis at higher orders.  This is based on exploiting geometrical connections that exist between the non-covered portion of the parameter space of a model, and correlations of the individual shapes $\{s_i\}$ in that model. For instance, in the extreme case where all $s_i$ have very large correlations with~$\T$, we are ensured that no order one linear combination of them can lie noticeably outside $\T$.  In Appendix \ref{geo} we study these geometric connections in more detail, and in Appendix \ref{templates} we use them to find new templates that must be added to $\T=\{{\rm Equil.,Ortho.}\}$ as we increase the number of derivatives one by one. Here, we only report the result: at 6-derivative level $\T=\{{\rm Equil.,Ortho.}\}$ is sufficient to cover the parameter space, at 7-derivative level we add one new template based on $\tdot\pi\pi_{,ij}^2$, at 8-derivative level the new 3-dimensional $\T$ is sufficient to approximately cover the parameter space, and finally, at 9-derivative level one new template based on the shape of $\tdot\pi\ddot\pi_{,i}^2$ is needed. Therefore, we propose the 4-dimensional $\T=\{{\rm Equil.,Ortho.},\tdot\pi\pi_{,ij}^2,\tdot\pi\ddot\pi_{,i}^2\}$ to search for inflationary models with up to 9-derivative interactions.  In the next section, we perform the optimal search for those templates in the WMAP 9 year data.

\section{Analysis of the WMAP data\label{sec:analysis}}

In the former section, we have shown that the parameter space for inflationary non-Gaussianities in single clock model 
should be enlarged to include two additional shapes. In the rest of the paper, we will precisely compute the additional
shapes and perform the optimal analysis of WMAP data to search for a signal.

For the sake of summary, we can say that the results of the former sections amount to us having to analyze the signatures of the following action:
\ba
S &=& \int d^3x \, dt \, \sqrt{-g} \, \Big( -M_{\rm Pl}^2 \dot H \Big) \Bigg[
   \frac{\dot\pi^2}{c_s^2} - \frac{(\partial_i\pi)^2}{a^2}  \nn \\ 
   &&\qquad \hspace{1cm}
   + {\mathcal C}_{\dot\pi^3} \frac{1}{c_s^2}  \dot\pi^3 
   + {\mathcal C}_{\dot\pi(\partial\pi)^2} \frac{1}{a^2} \dot\pi (\partial_i\pi)^2
\nn \\ 
&& \qquad\hspace{1cm}
   + {\mathcal C}_{\rm 7der} \frac{c_s^2}{a^4 H^4} \tdot\pi (\partial_i \partial_j \pi)^2
   + {\mathcal C}_{\rm 9der} \frac{1}{a^2 H^6} \tdot\pi \left( \partial_i \ddot\pi \right)^2
\Bigg]  \label{eq:action}\ ,
\ea
which contains the new 7-derivative and 9-derivative cubic operators
in addition to the 3-derivative operators $\dot\pi^3$ and $\dot\pi (\partial_i\pi)^2$.
We have written the coefficient of each operator as a dimensionless number ${\mathcal C}$,
multiplied by a combination of parameters $c_s,a,H$ chosen so that the three-point function
$\langle \zeta^3 \rangle$ at the end of inflation will be proportional to ${\mathcal C}$
with no dependence on $c_s$.
Note however that in single-clock models, the
coefficient ${\mathcal C}_{\dot\pi(\partial\pi)^2}$
is always related to the sound speed by ${\mathcal C}_{\dot\pi(\partial\pi)^2} = -(1-c_s^2)/c_s^2$.

\subsection{In-in calculation}

\par\noindent
In this section we will calculate the three-point function $\langle \zeta_{\k_1} \zeta_{\k_2} \zeta_{\k_3} \rangle$
for the action~(\ref{eq:action}) to lowest order in slow-roll parameters.
In this approximation the curvature perturbation at the end of inflation
is $\zeta = -H\pi$ and the primordial power spectrum is scale invariant.

The free QFT has the solution:
\ba
\pi_{\k}(\tau) 
  &=& \frac{H}{2 (-\dot H)^{1/2} M_{\rm Pl} c_s^{1/2} k^{3/2}} \left(  u_0(ic_sk\tau)^* a_{\k} + u_0(ic_sk\tau) a_{\k}^\dag \right) \nn \\
  &=& \frac{A_\zeta^{1/2}}{H k^{3/2}} \left(  u_0(i c_s k \tau)^* a_{\k} + u_0(i c_s k \tau) a_{\k}^\dag \right)\ ,
\ea
where $A_\zeta$ is the power spectrum amplitude defined by $P_\zeta(k) = A_\zeta / k^3$,
and the mode function $u_0(x)$ is:
\be
u_0(x) = (1 - x) e^x\ .
\ee
For taking proper (not conformal) time derivatives, it is useful to define
\be
u_n(x) = (-x \, d/dx)^n u_0(x)\ .
\ee
Then $(d/dt)^n u_0(i c_s k \tau) = H^n u_n(i c_s k \tau)$.
For reference, the first few $u_i$'s are:
\be
u_1(x) = x^2 e^x\ , \hspace{1cm}
u_2(x) = (-2x^2 - x^3) e^x\ , \hspace{1cm}
u_3(x) = (4x^2 + 5x^3 + x^4) e^x\ .
\ee
Now a long but straightforward calculation using the in-in formalism gives:
\be
\langle \zeta_{{\bf k}_1} \zeta_{{\bf k}_2} \zeta_{{\bf k}_3} \rangle 
  = \Big( 
      {\mathcal C}_{{\dot\pi}^3} F_{\dot\pi^3}
    + {\mathcal C}_{\dot\pi (\partial\pi)^2} F_{\dot\pi (\partial\pi)^2}
    + {\mathcal C}_7 F_7
    + {\mathcal C}_9 F_9
    \Big) (2\pi)^3 \delta^3(\k_1+\k_2+\k_3)\ ,
\ee
where the $F$'s are the following functions of momenta $k_i$:
\ba
F_{\dot\pi^3}(\k_1,\k_2,\k_3) &=& -\frac{3 A_\zeta^2}{k_1^3 k_2^3 k_3^3} \int_{-\infty}^0 \frac{d\tau_E}{\tau_E^4} u_1(k_1\tau_E) u_1(k_2\tau_E) u_1(k_3\tau_E) \nn\ , \\
F_{\dot\pi (\partial\pi)^2}(\k_1,\k_2,\k_3) &=& -A_\zeta^2 \frac{\k_2\cdot\k_3}{k_1^3 k_2^3 k_3^3}
     \int_{-\infty}^0 \frac{d\tau_E}{\tau_E^2} u_1(k_1\tau_E) u_0(k_2\tau_E) u_0(k_3\tau_E) + \mbox{2 perm.} \nn \ ,\\
F_{\rm 7der}(\k_1,\k_2,\k_3) &=& -A_\zeta^2 \frac{(\k_2\cdot\k_3)^2}{k_1^3k_2^3k_3^3} \int_{-\infty}^0 d\tau_E \, u_3(k_1\tau_E) u_0(k_2\tau_E) u_0(k_3\tau_E) + \mbox{2 perm.} \nn \ , \\
F_{\rm 9der}(\k_1,\k_2,\k_3) &=& -A_\zeta^2 \frac{\k_2\cdot\k_3}{k_1^3 k_2^3 k_3^3} \int_{-\infty}^0 \frac{d\tau_E}{\tau_E^2} u_3(k_1\tau_E) u_2(k_2\tau_E) u_2(k_3\tau_E) + \mbox{2 perm.}\ .
  \label{eq:in_in}
\ea
We have expressed each $F$ as an integral over a Euclidean conformal time $\tau_E$
defined by $\tau_E = ic_s\tau$. 
The integrals are straightforward to evaluate, but it will be convenient to leave them
unevaluated for reasons that will be apparent shortly.
The shapes of these four kinds of non-Gaussianities are given in Figure~\ref{fig:shapes}.

\begin{figure}[t]
\begin{center}
\includegraphics[width=7 cm]{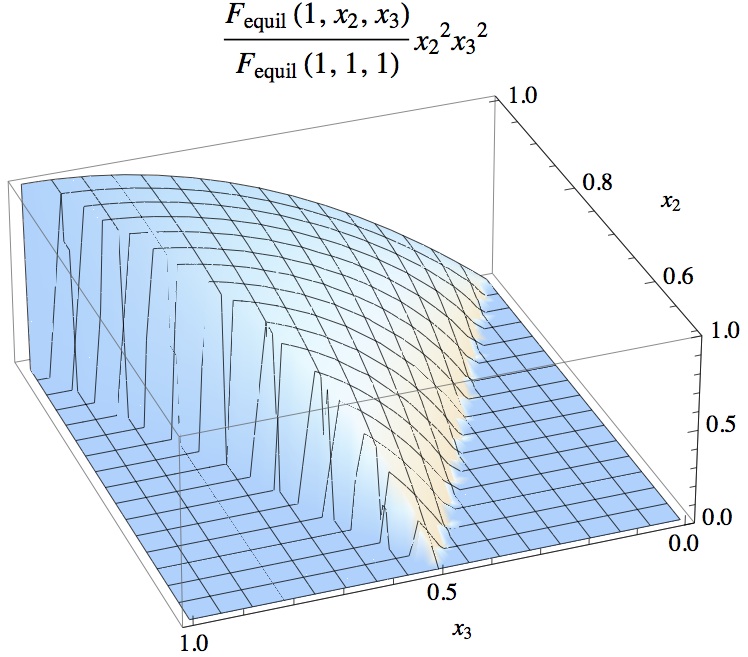}
\includegraphics[width=7 cm]{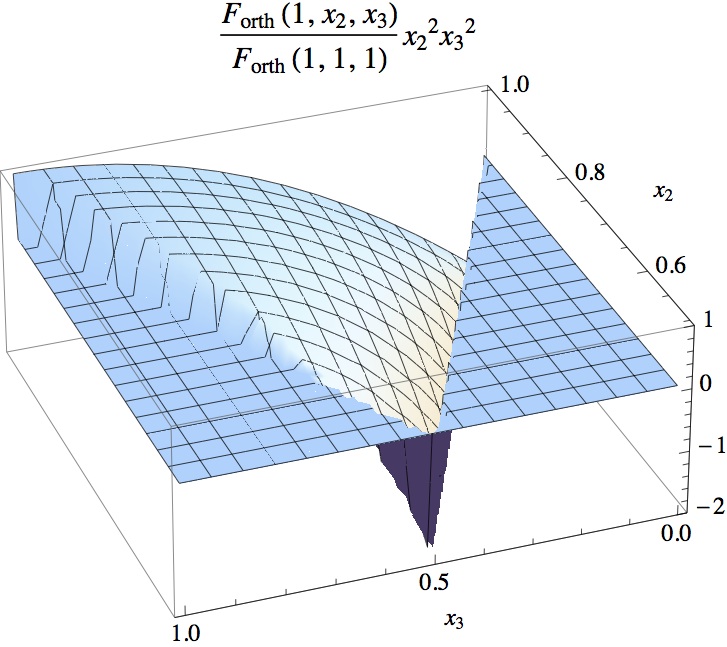}
\includegraphics[width=7 cm]{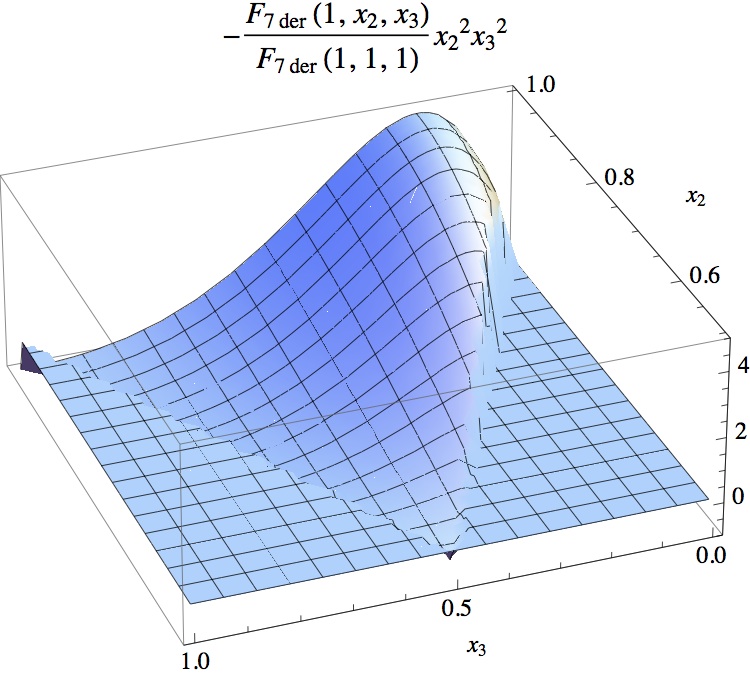}
\includegraphics[width=7 cm]{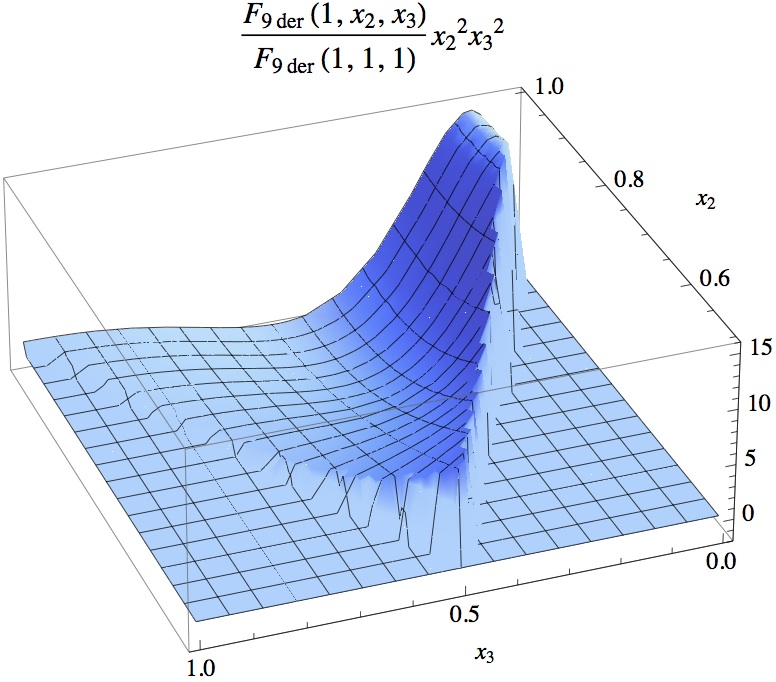}
\end{center}
\caption[]{The four shapes of non-Gaussianities from single-clock inflation.}
\label{fig:shapes}
\end{figure}

\subsection{Making the shapes factorizable} 

Our WMAP analysis procedure will follow the framework of~\cite{Smith:2006ud}
(see also~\cite{Wang:1999vf,Komatsu:2003iq,Creminelli:2005hu,Creminelli:2006rz,Yadav:2007ny,Senatore:2009gt}).
We briefly outline the key steps, referring to~\cite{Smith:2006ud} for details.

The first step is to approximate each bispectrum shape $F(k_1,k_2,k_3)$ being analyzed
by a sum of terms which are ``factorizable'', in the sense that
each term takes the form $f(k_1) g(k_2) h(k_3) + \mbox{perm.}$.
This factorizability condition is needed for two reasons: to apply a fast algorithm 
for calculating the angular CMB bispectrum $\langle a_{\ell_1m_1} a_{\ell_2m_2} a_{\ell_3m_3} \rangle$
from the initial curvature bispectrum $F$, and to apply a fast estimator for the bispectrum
amplitude given CMB data.

Several approaches have been proposed for finding factorizable approximations to a given
shape $F(k_1,k_2,k_3)$. For the 3-derivative shapes $(\dot\pi^3)$ and $\dot\pi(\partial\pi)^2$,
it is standard practice to approximate each shape by a linear combination of
the ``equilateral''~\cite{Creminelli:2005hu} and ``orthogonal''~\cite{Senatore:2009gt} templates:
\ba
F_{\rm eq}(k_1,k_2,k_3) &=& \frac{3}{5} A_\zeta^2 \left( \frac{6}{k_1^3 k_2^2 k_3} - \frac{3}{k_1^3 k_2^3} - \frac{2}{k_1^2 k_2^2 k_3^2} \right) + \mbox{5 perm.} \nn \ , \\
F_{\rm orth}(k_1,k_2,k_3) &=& \frac{3}{5} A_\zeta^2 \left( \frac{18}{k_1^3 k_2^2 k_3} - \frac{9}{k_1^3 k_2^3} - \frac{8}{k_1^2 k_2^2 k_3^2} \right) + \mbox{5 perm.}  \label{eq:templates}\ .
\ea
More precisely, we approximate the three-point function by:
\be
\langle \zeta_{{\bf k}_1} \zeta_{{\bf k}_2} \zeta_{{\bf k}_3} \rangle = \Big(
   f_{NL}^{\rm eq} F_{\rm eq}(k_1,k_2,k_3)
 + f_{NL}^{\rm orth} F_{\rm orth}(k_1,k_2,k_3)
\Big) (2\pi)^3 \delta^3(\k_1+\k_2+\k_3)  \label{eq:template_approximation}\ ,
\ee
where $\fnl^{\rm eq}$ and $\fnl^{\rm orth}$ are related to the coefficients in the action by:
\be
\left( \begin{array}{c}
  \fnl^{\rm eq}  \\
  \fnl^{\rm orth}
\end{array} \right) =
\left( \begin{array}{cc}
-0.0785 & 0.276 \\
0.0163 & -0.0157 
\end{array}
\right)
\left(
\begin{array}{c}
  {\mathcal C}_{\dot\pi^3} \\
  {\mathcal C}_{\dot\pi(\partial\pi)^2}
\end{array} \right)\ .
\ee
Although the equilateral and orthogonal templates are not precisely equal to the
bispectra $F_{\dot\pi^3}$ and $F_{\dot\pi(\partial\pi)^2}$,
the template approximation in Eq.~(\ref{eq:template_approximation}) has been shown to
be accurate to $\approx 99$\% in the case where $\fnl^{\rm eq} \gg \fnl^{\rm orth}$
and to $\approx 90$\% when $\fnl^{\rm eq} \ll \fnl^{\rm orth}$~\footnote{A  more precise template for  $\fnl^{\rm orth}$ was provided in the appendix of~\cite{Senatore:2009gt}, whose use, given absence of detection, has not so far been needed in CMB studies. This template is more accurate in the squeezed limit, and it should be the one to use for studies of scale dependent bias in large scale structures.}.

One minor technical point.
So far we have assumed scale invariance, when writing down the equilateral and orthogonal templates
in Eq.~(\ref{eq:templates}) and when doing the in-in calculations in Eq.~(\ref{eq:in_in}).
In the analysis of WMAP data, we will use the following slight modification of these shapes:
\ba
F_{\rm eq}(k_1,k_2,k_3) &=& \frac{3}{5} \bigg( 
    6 P_\zeta(k_1) P_\zeta(k_2)^{2/3} P_\zeta(k_3)^{1/3} 
  - 3 P_\zeta(k_1) P_\zeta(k_2) \nn \\
&& \hspace{1cm}
  - 2 P_\zeta(k_1)^{2/3} P_\zeta(k_2)^{2/3} P_\zeta(k_3)^{2/3} \bigg) + \mbox{5 perm.} \nn \ , \\
F_{\rm orth}(k_1,k_2,k_3) &=& \frac{3}{5} \bigg( 
    18 P_\zeta(k_1) P_\zeta(k_2)^{2/3} P_\zeta(k_3)^{1/3} 
  - 9 P_\zeta(k_1) P_\zeta(k_2) \nn \\
&& \hspace{1cm}
  - 8 P_\zeta(k_1)^{2/3} P_\zeta(k_2)^{2/3} P_\zeta(k_3)^{2/3} \bigg) + \mbox{5 perm.} \nn \ , \\
F_{\rm 7der}(\k_1,\k_2,\k_3) &=&
-\frac{(\k_2\cdot\k_3)^2}{k_1 k_2 k_3} P_\zeta(k_1)^{2/3} P_\zeta(k_2)^{2/3} P_\zeta(k_3)^{2/3} \nn  \\
  && \hspace{1cm}
\times \int_{-\infty}^0 d\tau_E \, u_3(k_1\tau_E) u_0(k_2\tau_E) u_0(k_3\tau_E) + \mbox{2 perm.} \nn \ , \\
F_{\rm 9der}(\k_1,\k_2,\k_3) &=&
-\frac{\k_2\cdot\k_3}{k_1 k_2 k_3} P_\zeta(k_1)^{2/3} P_\zeta(k_2)^{2/3} P_\zeta(k_3)^{2/3} \nn \\
  && \hspace{1cm} \times  \int_{-\infty}^0 \frac{d\tau_E}{\tau_E^2} u_3(k_1\tau_E) u_2(k_2\tau_E) u_2(k_3\tau_E) + \mbox{2 perm.}\ .  \label{eq:ns}
\ea
These definitions reduce to the previous one in the scale-invariant case $P_\zeta(k) = A_\zeta / k^3$, but
make sense if $P_\zeta(k)$ deviates slightly from scale invariance.

To represent the 7-derivative and 9-derivative shapes in factorizable form, rather than using templates, we use a physical approach
based on the observation that the in-in formalism automatically represents each shape as a conformal
time integral with factorizable integrand.
For each shape on the RHS of Eq.~(\ref{eq:in_in}), we first replace each factor of the form $(\k_2\cdot\k_3)$ by
\be
\k_2\cdot\k_3 \rightarrow \frac{1}{2}(k_1^2 - k_2^2 - k_3^2)\ ,
\ee
and then replace the $\tau_E$ integral by a finite sum which approximates it.
This procedure formally represents the shape $F(k_1,k_2,k_3)$ as a sum of factorizable terms; the key issue is whether
the number of terms needed to obtain an accurate approximation to the integral is manageably small.
We discretize the $\tau_E$ integrals using linear spacing in $\log|\tau_E|$
with 5 sampling points per decade, starting at $\tau_{E\rm min} = -10^6$~Mpc
and ending at $\tau_{E\rm max} = -0.04$~Mpc, for a total of 38 sampling points.
We then take the resulting factorizable approximation to $F(k_1,k_2,k_3)$
and compute the angular CMB bispectrum $\langle a_{\ell_1m_1} a_{\ell_2m_2} a_{\ell_3m_3} \rangle$.
To show that this discretization of the $\tau_E$ integrals has converged,
we do the following end-to-end test.
We recompute the angular CMB bispectrum using a coarser $\tau_E$ sampling, larger
$\tau_{E\rm min}$, and smaller $\tau_{E\rm max}$.
We then verify that the two CMB bispectra agree (using the metric defined by the Fisher
matrix, which corresponds to observational distinguishability) at the $\approx 10^{-5}$ level.
We have done an analogous convergence test for other numerical parameters involved in the
bispectrum calculation: the CAMB~\cite{Lewis:1999bs} accuracy settings used to compute the CMB line-of-sight source function;
the spacing in the time integral used to compute the CMB transfer function $\Delta_\ell(k)$ from the source function; and
the endpoints/spacing of the $k$-integral and $r$-integrals used to compute the CMB angular bispectrum from the transfer function.
Taken together, these tests show that we have obtained factorizable representations for the 7-derivative and 9-derivative shapes
which allow the CMB bispectra to be approximated with negligibly small residual.

The above procedure represents the CMB bispectrum as a sum of terms which obey an angular
factorizability condition.
This representation contains a large number of terms but is redundant: most terms can be
approximated as linear combinations of a small subset of ``independent'' terms.
The optimization algorithm from~\cite{Smith:2006ud} takes advantage of this redundancy
to produce a more efficient factorizable representation while ensuring that
the bispectrum is unchanged within a small numerical threshold.
In Table~\ref{tab:nfact}, we show the improvement in the number of factorizable terms
$N_{\rm fact}$ which results from the optimization algorithm.
After optimization, the 7-derivative and 9-derivative shapes have factorizable representations
with $N_{\rm fact}$ of order a few hundred, which is small enough for practical data analysis.
Furthermore, every step of the procedure used to obtain these factorizable representations
is a controlled approximation in which the residual is guaranteed to be small~\footnote{We note that the same procedure could have been
applied to the $\dot\pi^3$ and $\dot\pi(\partial\pi)^2$ shapes, using the integral representations
in Eq.~(\ref{eq:in_in}) obtained from the in-in formalism.  We have not done this in order to
facilitate comparison with previous results, which use the equilateral and orthogonal template
approximations.  As previously remarked, these approximations are accurate at the $\approx$99\%
and $\approx$90\% level, so the difference is not very important for practical purposes.}.

\begin{table}
\begin{center}
\begin{tabular}{|c||c|c|} \hline
Shape & $N_{\rm fact}$ (pre-optimization) & $N_{\rm fact}$ (post-optimization)  \\ \hline\hline
Equilateral template & 1326 & 93 \\ \hline
Orthogonal template & 1326 & 120 \\ \hline
7-derivative shape & 126396 & 388 \\ \hline
9-derivative shape & 63198 & 222 \\ \hline
\end{tabular}
\end{center}
\caption{Number of terms $N_{\rm fact}$ in the factorizable representations for the angular
CMB bispectrum $\langle a_{\ell_1m_1} a_{\ell_2m_2} a_{\ell_3m_3} \rangle$ for the four shapes
analyzed in this paper.}
\label{tab:nfact}
\end{table}

\subsection{WMAP results and interpretation}

We analyze WMAP data using the same pipeline and analysis parameters as in the WMAP9 results paper~\cite{Bennett:2012zja}.
This pipeline optimally combines data from $V$-band and $W$-band channels using inverse covariance weighting,
removes regions of large foreground contamination using the KQ75 mask, and projects out residual foregrounds
by marginalizing spatial templates for synchrotron, free-free and dust emission.
For more details of the pipeline, see \S8.1 of~\cite{Bennett:2012zja}.

We will analyze the equilateral, orthogonal, 7-derivative, and 9-derivative shapes.
The equilateral and orthogonal shapes have already been analyzed on the same data set in~\cite{Bennett:2012zja}, 
but we include them here since we will consider parameter spaces which include equilateral
and orthogonal non-Gaussianity in addition to our new shapes.

For historical reasons, it is conventional to normalize bispectrum coefficients by defining
$\fnl$ parameters so that $\langle \zeta_{\k_1} \zeta_{\k_2} \zeta_{\k_3} \rangle = (18/5) \fnl P_\zeta(k)^2$
on equilateral triangles satifying $k_1=k_2=k_3=k$.
The equilateral and orthogonal shapes have been defined in Eq.~(\ref{eq:ns}) so that they have this normalization.
To normalize our new shapes, we first
evaluate Eq.~(\ref{eq:ns}) on equilateral triangles, obtaining:
\be
F_{\rm 7der}(k,k,k) = \frac{17}{162} P_\zeta(k)^2\ , \hspace{1.5cm}
F_{\rm 9der}(k,k,k) = \frac{8}{729} P_\zeta(k)^2\ ,
\ee
so we define $\fnl$ parameters in terms of coefficients of the action~(\ref{eq:action}) by
\be
\fnl^{\rm 7der} = \frac{85}{2916} {\mathcal C}_7 \ ,\hspace{1.5cm} \fnl^{\rm 9der} = \frac{20}{6561} {\mathcal C}_9\ .
\ee
Using these definitions and the pipeline described above, our ``bottom line'' WMAP constraints on $\fnl$ parameters are:
\ba
f_{NL}^{\rm eq} &=& 51 \pm 136 \hspace{1.94cm} \mbox{($-221 < f_{NL}^{\rm eq} < 323$ at 95\% CL)}  \nn \\
f_{NL}^{\rm orth} &=& -245 \pm 100 \hspace{1.5cm} \mbox{($-445 < f_{NL}^{\rm orth} < -45$ at 95\% CL)}  \nn \\
f_{NL}^{\rm 7der} &=& -34 \pm 56 \hspace{1.85cm} \mbox{($-146 < f_{NL}^{\rm 7der} < 78$ at 95\% CL)}  \nn \\
f_{NL}^{\rm 9der} &=& 30 \pm 16 \hspace{2.12cm} \mbox{($-1 < f_{NL}^{\rm 9der} < 62$ at 95\% CL)}  \label{eq:bottom_line}
\ea
As reported in~\cite{Bennett:2012zja}, there is a $2.5\sigma$ preference for nonzero $f_{NL}^{\rm orth}$.
Note that our estimates of $\fnl^{\rm eq}$ and $\fnl^{\rm orth}$ agree perfectly with~\cite{Bennett:2012zja},
since the maps and pipeline are identical.

Each ``bottom line'' constraint in Eq.~(\ref{eq:bottom_line}) is actually the value of an estimator $\hfnl$
which is constructed assuming that the other three shapes are absent.  To analyze multiparameter spaces,
we need the correlation matrix of the four $\hfnl$ estimators, which we find using the WMAP pipeline to be:
\be
\left( \begin{array}{cccc}
  1 & 0.29 & -0.27 & 0.12 \\
 0.29 & 1 & 0.51 & -0.73 \\
-0.27 & 0.51 & 1 & -0.58 \\
 0.12 & -0.73 & -0.58 & 1 
\end{array} \right)  \label{eq:corr}\ .
\ee
We can quantify the total deviation of all four $\fnl$ parameters from zero by computing
\be
\chi^2 = (\hfnl)^T C^{-1} (\hfnl)  \label{eq:chi2}\ .
\ee
Here, $(\hfnl)$ is a vector containing the four best-fit values in Eq.~(\ref{eq:bottom_line}), 
$C_{ij} = r_{ij} \sigma_i \sigma_j$ is the covariance matrix obtained by combining the $1\sigma$ errors $\sigma_i$ in Eq.~(\ref{eq:bottom_line})
and the correlation matrix $r_{ij}$ in Eq.~(\ref{eq:corr}).
For a partial parameter space with $N < 4$ shapes, we compute $\chi^2$ by reducing $C$ to an $N$-by-$N$
matrix (by removing rows and columns) before taking the matrix inverse in Eq.~(\ref{eq:chi2}). 

\begin{table}
\begin{center}
\begin{tabular}{|c||c|c|} \hline
  Parameter space & $\chi^2 / \mbox{(d.o.f.)}$ & $p$-value  \\ \hline\hline
  $\{$Equil$\}$  &  0.14 / 1 & 0.71 \\
  $\{$Equil, Orth$\}$  &  7.3 / 2 & 0.026 \\
  $\{$Equil, Orth, 7der$\}$  &  9.7 / 3 & 0.022 \\
  $\{$Equil, Orth, 7der, 9der$\}$  &  9.7 / 4 & 0.046 \\ \hline
\end{tabular}
\end{center}
\caption{Statistical siginificance of deviation from Gaussian statistics,
as quantified by the $\chi^2$ statistic in Eq.~(\ref{eq:chi2}), for a sequence of
parameter spaces obtained by sequentially adding shapes with increasing numbers
of derivatives.}
\label{tab:chi2}
\end{table}

In Table~\ref{tab:chi2}, we show $\chi^2$ values for a sequence of parameter spaces
obtained by sequentially adding shapes with increasing numbers of derivatives.
The $p$-value associated with each $\chi^2$ is the probability that a Gaussian
simulation will give a $\chi^2$ larger than the WMAP data.
The small $p$-value in the second row of the table is driven by the preference
for nonzero $\fnl^{\rm orth}$ in WMAP9.
It is interesting to observe that the $p$-value in the third row is slightly lower,
i.e.~adding the 7-derivative shape to the $\{$Equil, Orth$\}$ parameter space
slightly increases the statistical evidence for non-Gaussianity.
However, the two $p$-values are similar enough that our interpretation of Table~\ref{tab:chi2}
is that the marginal evidence for non-Gaussianity in WMAP is mainly driven by $\fnl^{\rm orth}$.

\begin{figure}
\centerline{\includegraphics[width=12cm]{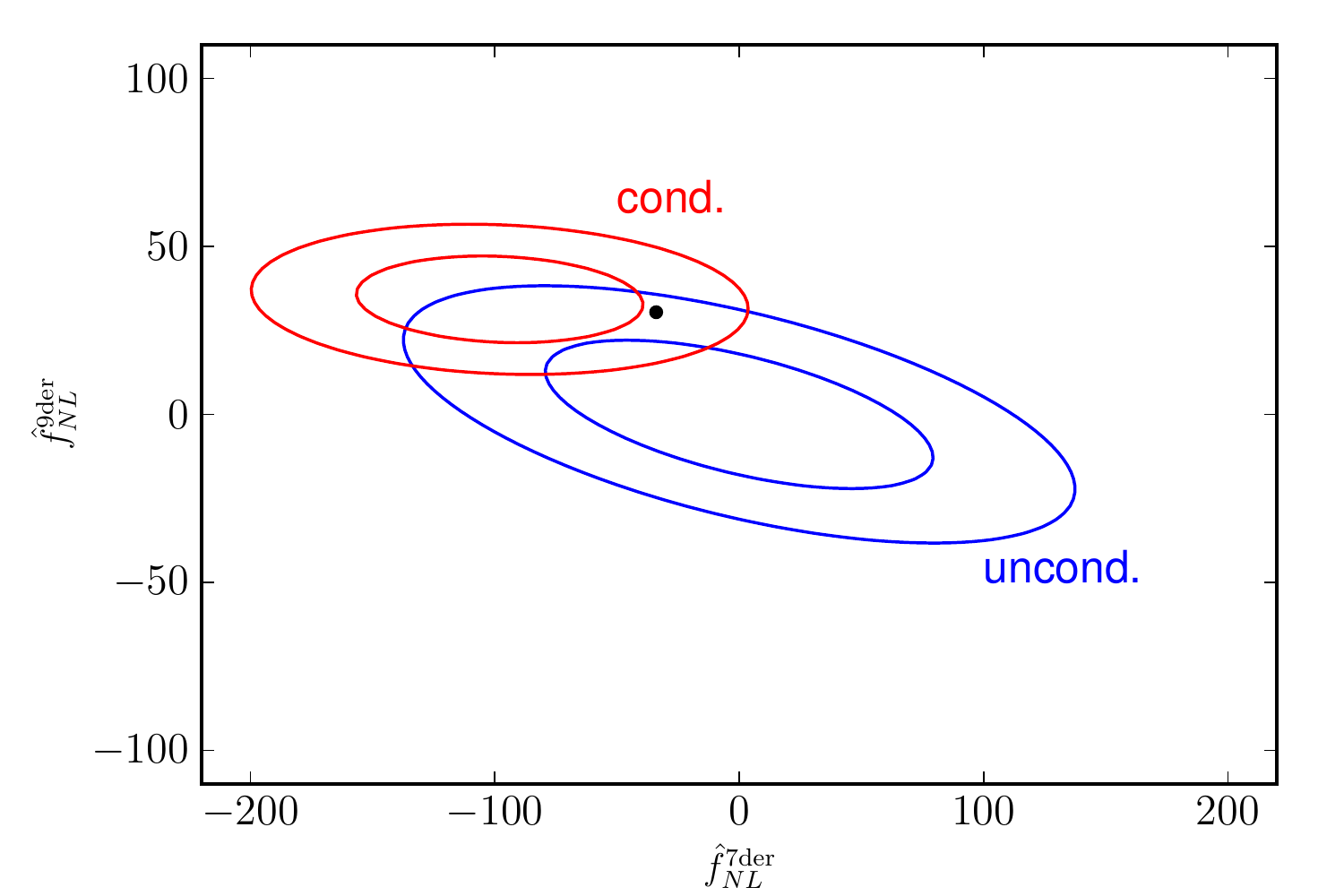}}
\caption{Error ellipses in the $(\fnl^{\rm 7der}, \fnl^{\rm 9der})$-plane,
with WMAP values shown (black point).
The ellipses labeled ``uncond.'' are 68\% and 95\% confidence regions 
obtained from an ensemble of Gaussian simulations, and the ellipses labeled ``cond.''
are confidence regions obtained by postselecting only those simulations whose
values of $(\fnl^{\rm eq}, \fnl^{\rm orth})$ agree with the WMAP values.}
\label{fig:ellipses}
\end{figure}

There is one counterintuitive aspect of this table which deserves further comment.
The ``bottom line'' result in Eq.~(\ref{eq:bottom_line}) suggests that there is weaker
evidence for $\fnl^{\rm 7der}$ than $\fnl^{\rm 9der}$ ($0.6\sigma$ versus $2\sigma$),
whereas Table~\ref{tab:chi2} suggests the opposite ($\Delta\chi^2 = 2.4$
when the 7-derivative shape is added, versus $\Delta\chi^2 < 0.1$ for the 9-derivative shape).
This can be understood as follows.
In Fig.~\ref{fig:ellipses} we show two sets of error ellipses in the 
$(\fnl^{\rm 7der}, \fnl^{\rm 9der})$ plane.
The first set (``uncond'') represents confidence regions that would be obtained from an ensemble 
of Gaussian simulations, and the second set (``cond.'') represents confidence regions that would
be obtained by postselecting only those simulations whose values of $(\fnl^{\rm eq}, \fnl^{\rm orth})$
lie within narrow intervals centered on the WMAP values $(\fnl^{\rm eq}, \fnl^{\rm orth}) = (51, -245)$.
The two sets of ellipses differ significantly because $\fnl^{\rm orth}$ is nonzero at 2.5$\sigma$,
and the two higher-derivative shapes are significantly correlated with the orthogonal shape.
Depending on which set of ellipses one interprets the WMAP point $(\fnl^{\rm 7der}, \fnl^{\rm 9der}) = (-34,30)$
relative to, either $\fnl^{\rm 7der}$ or $\fnl^{\rm 9der}$ may appear to be more anomalous.
This explains the apparent discrepancy between Eq.~(\ref{eq:bottom_line}), where each shape is
estimated assuming the other three shapes are zero, and Table~\ref{tab:chi2}, where the total
statistical evidence for non-Gaussianity is accumulated accounting for correlations between shapes.

As mentioned in the introduction, we do not analyze Planck data since Monte Carlo simulations of the foreground-cleaned Planck maps
are not yet public, and would be impractical to construct due to complexity of the Planck noise model.
We note that the equilateral and orthogonal templates have been analyzed by the Planck collaboration in~\cite{Ade:2013ydc},
where it was found that the additional high-$\ell$ information degrades the 2.5$\sigma$ anomaly for $\fnl^{\rm orth}$ in WMAP9 data to $1.4\sigma$ in Planck:
\ba
f_{NL}^{\rm eq, \; \rm Planck} &=& 63 \pm 57 \hspace{1.94cm} \mbox{($-51 < f_{NL}^{\rm eq, \; \rm Planck} < 177$ at 95\% CL)}  \nn \\
f_{NL}^{\rm orth,\;\rm Planck} &=& -52 \pm 37 \hspace{1.5cm} \mbox{($-127 < f_{NL}^{\rm orth, \; \rm Planck} < 21$ at 95\% CL)}  \nn 
\ea
Given that the low $p$-value that we see in WMAP 9yr after including the 7- and 9-derivative shapes seems to be driven by the anomaly in the orthogonal shape, 
it will be interesting to see the results of searching for the 7- and 9-derivative shapes in Planck data.

\section*{Acknowledgments}

 We thank Andrei Gruzinov, Ami Katz, and Matias Zaldarriaga for useful discussions. The research of S.R.B is supported by the DOE under grant numbers DE-FG02-01ER-40676 and DE-FG02-01ER-40676. MM is supported by NSF Grant PHY-1314311.
L.S. is supported by DOE Early Career Award DE-FG02-12ER41854 and by NSF grant PHY-1068380. 
 Research at Perimeter Institute is supported by the Government of Canada
 through Industry Canada and by the Province of Ontario through the Ministry of Research \& Innovation.
 Some computations were performed on the GPC cluster at the SciNet HPC Consortium.
 SciNet is funded by the Canada Foundation for Innovation under the auspices of Compute Canada,
 the Government of Ontario, and the University of Toronto.
 KMS was supported by an NSERC Discovery Grant.

\appendix
\section{\label{natural} Technical Naturalness}

In this appendix we apply the non-renormalization theorem that we have explained in the main text to the context of the EFTofI, where, as explained below, additional constraints arise because of non-linear realization of Lorentz invariance. Let us consider the following high derivative inflationary model as a prototype and explain why it is a technically natural model:
\bea
S=\int d^4 \tilde x\; a^3 \left[\dot{\pi_c}^2-{\pi_c}_{,i}^2+\frac{H^6}{\Lambda^8}(a_3\dot\pi_c{\pi_c}_{,i}^2+a'_3\dot\pi_c^3)+\frac{1}{\Lambda^8}(a_9\tdot\pi_c^3+a_9' \tdot\pi_c {\pi_c}_{,ijk}^2)\right].
\label{S2}
\eea
where $a_{3,9}$ and  $a_{3,9}'$ are coefficients of order unity, and, $\pi$ is canonically normalized according to the prescription of footnote \ref{cs}~\footnote{When $c_s\lesssim 1$, we need to ensure that the induced  $\dot\pi\d_i\pi\d_i\pi$ operator induces a subleading level of non-Gaussianity. By the way we wrote (\ref{S2}), this is automatically enforced.}. The generic features of this model that would serve as a guide to build other technically natural models are the following:

{\bf i)} We have only considered high derivative interactions that are not connected to the quadratic action by time diffs. Otherwise there would be large high derivative corrections to the kinetic term, which, if time-like, would introduce a ghost in the theory, and we wish in general to keep the kinetic term unaffected~\footnote{We leave the extension of our analysis to the case of more complicated kinetic terms, such as the ghost-condensate-like, $\omega^2\sim k^4/M^2$, to future work.}. In practice this is insured by studying unitary gauge operators that start from cubic order in the perturbations. For instance, the unitary gauge action that gives rise to \eqref{S2}, apart from operators that are written in~\eqref{S}, must contain
\bea
\label{unit}
({\partial^0}^2g^{00})^3,\quad \text{and}\quad {\partial^0}^2g^{00}(\nabla_i\delta E_{jk})^2\,.
\eea

{\bf ii)} To realize time diffs. some of the high derivative cubic terms are accompanied by lower derivative ones multiplied by derivatives of the metric. They can be obtained by tedius expansions of unitary gauge operators like \eqref{unit} in terms of $\pi$ (for instance $\delta E_{ij}=-\pi_{,ij}-\dot g_{ij} \dot\pi/2 +\cdots$). If the metric is evaluated on the background, these extra terms are suppressed by powers of $H/\Lambda$ and, despite having less derivatives, do not dominate over the higher derivative terms we are interested in. Moreover, one can find linear combinations of unitary gauge operators with $\mathcal{O}(1)$ coefficients in which those companions are absent (e.g. $\delta E_{ij}+H g_{ij}{\delta g}^{00}/2$). 

{\bf iii)} The unitary gauge operators of interest always contain higher than cubic interactions of $\pi$ whose loops can renormalize the quadratic and cubic action (e.g. \eqref{unit} contains $\ddot\pi^2\pi_{,ijk}^2$). However, after canonically normalizing $\pi$, these higher order interactions are suppressed by extra factors of $1/(-\mpl^2\dot H c_s)^{1/2}$, and as a result, these loops are suppressed by powers of $\Lambda_{UV}^2/(-\mpl^2\dot H c_s)^{1/2}$, as observed in~\cite{4point}. Using the amplitude of power spectrum,~$\zeta$, introduced below eq. \eqref{fnl}, this ratio can be written as $\zeta\Lambda_{UV}^2/H^2\simeq 5\times 10^{-5}\Lambda_{UV}^2/H^2$. This is much less than one in models with large non-Gaussianity where the strong coupling scale is relatively close to $H$. We can therefore safely ignore these higher order terms. 

{\bf iv)} Consider a high derivative interaction $I$ and a lower derivative one $I'$. A simple generalization of argument in the main text for relativistic theories applies to the structure of the counter-terms in a non-relativistic Lagrangian: if the derivative structures of $I$ and $I'$ do not match, then $I$ will not renormalize $I'$. In other words, if there is no loop diagram of $I$ whose external legs are the same as the legs of $I'$, no counterterm proportional to~$I'$ is needed in renormalization of $I$. We thus conclude that the last interaction in \eqref{S2} will not modify the form of the lower derivative part of the action through loops. The coefficients $a_3,a_3'$ are, therefore, independent of $a_9,a_9'$ and can be set to zero.

{\bf v)} If there existed a loop of $I$ that matched $I'$ but $I$ and $I'$ had different parity under $t\to -t$, then $I'$ will be renormalized with a suppression of $H/\Lambda$. For instance $\ddot\pi_c\dot{\pi_c}_{,ij}\pi_{c,ij}/\Lambda^6$ naturally comes with $H\ddot\pi_{c}^3/\Lambda^6$. This can be understood by noticing that in this case $H$ is the symmetry breaking scale of the $t\to t'$ symmetry.  

In general, all technically natural models of single clock inflation can be written by using the above rules, which amounts to scanning all derivative interactions (or their combinations with the hierarchy given by \eqref{hierarchy}) and rejecting those which cannot be constructed out of unitary gauge operators, or get largely modified after renormalization. There is in fact an enormous number of such models since there are many cubic interactions with at most 9 derivatives. 


\section{\label{app:reduction} Classification of independent cubic operators}

We consider cubic operators $\O$ where at least one derivative (either time or spatial) acts on each of the three $\pi$'s,
so that the shift symmetry $\pi \rightarrow \pi + c$ is satisfied.
We can manipulate such operators using the classical equation of motion 
\be
\ddot\pi + 3 H \dot\pi - c_s^2 \partial^2 \pi = 0 \label{eq:eom}\ ,
\ee
which, as we discussed, is equivalent for our purposes to a field redefinition, or by adding a total derivative.
When we add a total derivative of the form $(\partial_t \O')$ or $(\partial_i \O'_i)$, the operator $\O'$ which appears
must satisfy the shift symmetry, i.e.~at least one derivative must act on each $\pi$.

We define operators
\ba
\O^{(1)}_{mnp} &=& (\partial_t^m \pi) (\partial_t^n \pi) (\partial_t^p \pi) \hspace{0.74cm} \mbox{(where $m \ge n \ge p \ge 1$)} \label{eq:O1} \ , \\
\O^{(2)}_{mn} &=& (\partial_i \pi) (\partial_i \partial_t^m \pi) (\partial_t^n \pi) \hspace{0.5cm} \mbox{(where $m \ge 0$ and $n \ge 1$)} \ .\label{eq:O2}
\ea
The goal of this appendix is to prove that {\em every shift-symmetric cubic operator $\O$ is a linear combination
of operators of the form $\O^{(1)}_{nnp}$ and $\O^{(2)}_{0,2n-1}$.}  The proof will be split into four steps as follows.

First we prove the weaker statement that every operator is a linear combination
of operators $\O^{(1)}_{mnp}$ and $\O^{(2)}_{mn}$.
We prove this by induction on the number $S$ of spatial derivatives.
If $S=0$, then $\O$ is already of the form $\O^{(1)}_{mnp}$, so we can assume $S \ge 1$.
We can assume that $\partial_i^2$ does not act on any of the $\pi$'s, since we
can use the equation of motion~(\ref{eq:eom}) to replace $\partial^2\pi$ by a linear combination of $\ddot\pi, \dot\pi$, 
decreasing $S$.
Thus we can assume the operator is of the form
\be
\O = (\partial_i \D \pi) (\partial_i \D' \pi) (\D'' \pi)  \label{eq:proof1}\ ,
\ee
where $\D,\D',\D''$ are differential operators.
If $\D \ne 1$ and $\D'\ne 1$, we write
\be
\O = \frac{1}{2} \partial^2 ( \D\pi \D'\pi ) (\D''\pi) - (\partial^2 \D\pi) (\D'\pi) (\D''\pi) - (\D\pi) (\partial^2 \D'\pi) (\D''\pi)\ ,
\ee
and then integrate by parts in the first term to put the $\partial^2$ on $\D''\pi$.  Then we can use the equation of motion to
replace $\partial^2$ by spatial derivatives and decrease $S$.  
Thus we can assume that either $\D=1$ or $\D'=1$ in Eq.~(\ref{eq:proof1}), i.e. $\O$ is of the form
\be
\O = (\partial_i \pi) (\partial_i \D'\pi) (\D''\pi)   \label{eq:proof2}\ .
\ee
If $\D'$ and $\D''$ contain no spatial derivatives, then $\O$ is of the form $\O^{(2)}_{mn}$.
Otherwise, $\D'$ and $\D''$ must each contain a factor $\partial_j$, and
the same argument that led to the form~(\ref{eq:proof2}) shows that the most general possibility is the operator
\be
\O = (\partial_i \pi) (\partial_i \partial_j \partial_t^n \pi) (\partial_j \pi) \hspace{1cm} (n \ge 0)   \label{eq:proof_n}\ .
\ee
Up to total derivative terms 
\bea
\O = - (\d^2\pi)(\d_j\d_t^n\pi) (\d_j\pi) - \frac{1}{2}\d_j(\d_i\pi\d_i\pi)(\d_j\d_t^n\pi) \nonumber\\
= - (\d^2\pi)(\d_j\d_t^n\pi)(\d_j\pi) + \frac{1}{2}\d_i\pi\d_i\pi(\d^2\d_t^n\pi)\ .
\eea
Both terms on the r.h.s. reduce to $\O^{(2)}$ operators when the equation of motion is used. This completes the inductive proof that any $\O$ is a linear combination of $\O^{(1)}_{mnp}$ and $\O^{(2)}_{mn}$.

Second, we prove that $\O^{(1)}_{mnp}$ is a linear
combination of operators of the form $\O^{(1)}_{nnp}$.
We prove this by induction on $m$.
If $m \ge n+2$ then we can integrate by parts to decrease $m$.
If $m = n+1$ and $n > p$ then we can write
\ba
f(t)\;\O &=& f(t)\; \frac{1}{2} \partial_t \Big( (\partial_t^n \pi) (\partial_t^n \pi) \Big) (\partial_t^p \pi) \nn \\
  & \to & -f(t)\; \frac{1}{2} (\partial_t^n \pi) (\partial_t^n \pi) (\partial_t^{p+1} \pi)-f'(t)\;  \frac{1}{2} (\partial_t^n \pi) (\partial_t^n \pi) (\partial_t^{p} \pi)\ ,
\ea
which is of the desired form.
If $m = n+1$ and $n=p$ then we can write
\be
f(t)\; \O =f(t)\; \frac{1}{3} \partial_t \Big( (\partial_t^n \pi) (\partial_t^n \pi) (\partial_t^n \pi) \Big) =-f'(t) \;\frac{1}{3}  (\partial_t^n \pi) (\partial_t^n \pi) (\partial_t^n \pi)\ ,
\ee
which is of the desired form.
If $m=n$, then $\O$ is already in the desired form $\O^{(1)}_{nnp}$.
This covers all cases and completes the proof that $\O^{(1)}_{mnp}$ is a linear combination of operators of the form $\O^{(1)}_{nnp}$.

Third, we prove that $\O^{(2)}_{mn}$ is reducible to $\O^{(2)}_{0,p}$ operators with $p \leq m+n$. 
We introduce the notation $\sim$ to mean ``equal up to integration by parts and terms of the form $\O^{(1)}_{mnp}$''. 
For $m \ge 1$ and $n \ge 1$, we have:
\ba
f(t)\;\O^{(2)}_{1n} 
  & = &f(t)\; \frac{1}{2} \partial_t \Big( (\partial_i\pi) (\partial_i\pi) \Big) (\partial_t^n\pi) \nn \\
  & \to & -f(t)\;\frac{1}{2} (\partial_i \pi) (\partial_i \pi) (\partial_t^{n+1}\pi)-f'(t)\;\frac{1}{2} (\partial_i \pi) (\partial_i \pi) (\partial_t^{n}\pi) \nn \\
  & = & -\frac{1}{2}f(t) \O^{(2)}_{0,n+1}  -\frac{1}{2}f'(t) \O^{(2)}_{0,n}
\ea
Similarly
\ba
f(t)\;\O^{(2)}_{mn} \sim -f(t)\; \O^{(2)}_{m-1,n+1} - f'(t)\;\O^{(2)}_{m-1,n}\quad \text{for }\quad m>1.
\ea
By iterating we get the desired result.

Finally, we show that $\O^{(2)}_{0,2n}$ are reducible to  combinations of $\O^{(2)}_{0,2p+1}$ with $2p+1<2n$ and $\O^{(1)}$. Suppose all $\O^{(2)}_{0,2m}$ with $m<n$ are reducible. By integrating by parts, we have
\ba
f(t) \O^{(2)}_{0,2n}\sim (-1)^n  2 f(t)\O^{(2)}_{nn}+ \text{linear combination of $\O^{(2)}_{pq}$ with $p+q<2n$.}
\ea
Using the third step above, the second class of terms can be reduced to $\O^{(1)}$'s and $\O^{(2)}_{0,N}$ with $N<2n$, which by  induction can only contain odd $N$. 
The first term  can be reduced to $\O^{(1)}$'s by noting that
\ba
f(t) \O^{(2)}_{nn} 
  &=& f(t) (\partial_i \pi) (\partial_i \partial_t^n \pi) (\partial_t^n \pi) \nn \\
  &=& \frac{1}{2} f(t) (\partial_i \pi) \partial_i \Big( (\partial_t^n\pi) (\partial_t^n\pi) \Big) \nn \\
  & \to & -\frac{1}{2} f(t) (\partial^2 \pi) (\partial_t^n\pi) (\partial_t^n\pi)
\ea
and using the equation of motion~(\ref{eq:eom}) to trade $(\partial^2\pi)$ for time derivatives.
This completes the proof.

So far, we have neglected cubic-in-$\pi$ operators that contain the $\epsilon^{ijk}$ symbol to contract some indexes. It is easy to check that these operators vanish. In fact, with only one $\epsilon^{ijk}$, we can have
\be
f(t)\; \epsilon^{ijk} (\partial_i \D \pi) (\partial_j \D' \pi) (\partial_k \D'' \pi)\ ,
\ee
where $\D,\,\D',\,\D''$ are allowed to have time derivatives or spatial derivatives which contract with each other. After we integrate by parts any of the spatial derivatives, the operator vanishes. The same holds for any number of $\epsilon^{ijk}$'s we insert: they cannot be contracted among themselves, as otherwise reduce to a sum of products of Kronecker-$\delta$'s, decreasing the number of $\epsilon^{ijk}$'s. Therefore each index in $\epsilon^{ijk}$ need to be contracted with a spatial derivative acting on each of the three different $\pi$'s. Since we have only three $\pi$'s, the operator is zero after integrating by parts one of the spatial derivatives. This exhausts all possible operators, as we wanted to show.

\section{\label{geo} Non-covered parameter space from a geometrical viewpoint}

In this appendix we describe a set of connections between the portion of the parameter space of an inflationary model that is not covered by a template basis $\T$, and correlations of individual shapes appearing in that model among themselves and with $\T$.

Let us start with the simplest case where $\T$ is one dimensional, $\T =\{t_1\}$, and we have an inflationary model with two interactions. The shapes of these interactions form a two-dimensional set $\SS=\{s_1,s_2\}$, where $s_{1,2}$ are assumed to be normalized. Let us further assume that $s_1=t_1$; that is, $s_1$ is completely parallel to $\T$, which together with the normalization assumption implies ${s_1}_\parallel=1$ (the perpendicular and parallel components of a shape with respect to $\T$ are defined below \eqref{theta}).  Assuming that ${s_2}_{\perp}\neq 0$, we want to assess the portion of parameter space which is not covered by the one dimensional $\T$. Here, the parameter space is the two dimensional space $(\alpha_1,\alpha_2)$ of the coefficients of $s_\alpha=\alpha_1s_1+\alpha_2 s_2$, where $\alpha_{1,2}$ are of order unity. Taking as criterion for a shape $s_\alpha$ not being covered by $\T$ that $\cos(s_\alpha,\T)<0.7$, or $|{s_\alpha}_\perp|>|{s_\alpha}_\parallel|$, we get the condition
\bea
\label{coverage}
|\alpha_2 {s_2}_\perp|>|\alpha_1+\alpha_2 {s_2}_\parallel |\,,
\eea
which is a wedge between the lines $\alpha_1=({s_2}_\perp-{s_2}_\parallel)\alpha_2$ and $\alpha_1=-({s_2}_\perp+{s_2}_\parallel)\alpha_2$ in $\alpha$ space (see Fig. \ref{a1_a2} for the particular case where $\sin\theta_{s_2,\T}=0.8$). As a result, at fixed $\alpha_2$, $\alpha_1$ varies in a range of size
\bea
\label{delta_alpha1}
\Delta \alpha_1= 2 {s_2}_\perp \alpha_2.
\eea
\begin{figure}[t]
\begin{center}
\includegraphics[width=10 cm, height= 8 cm]{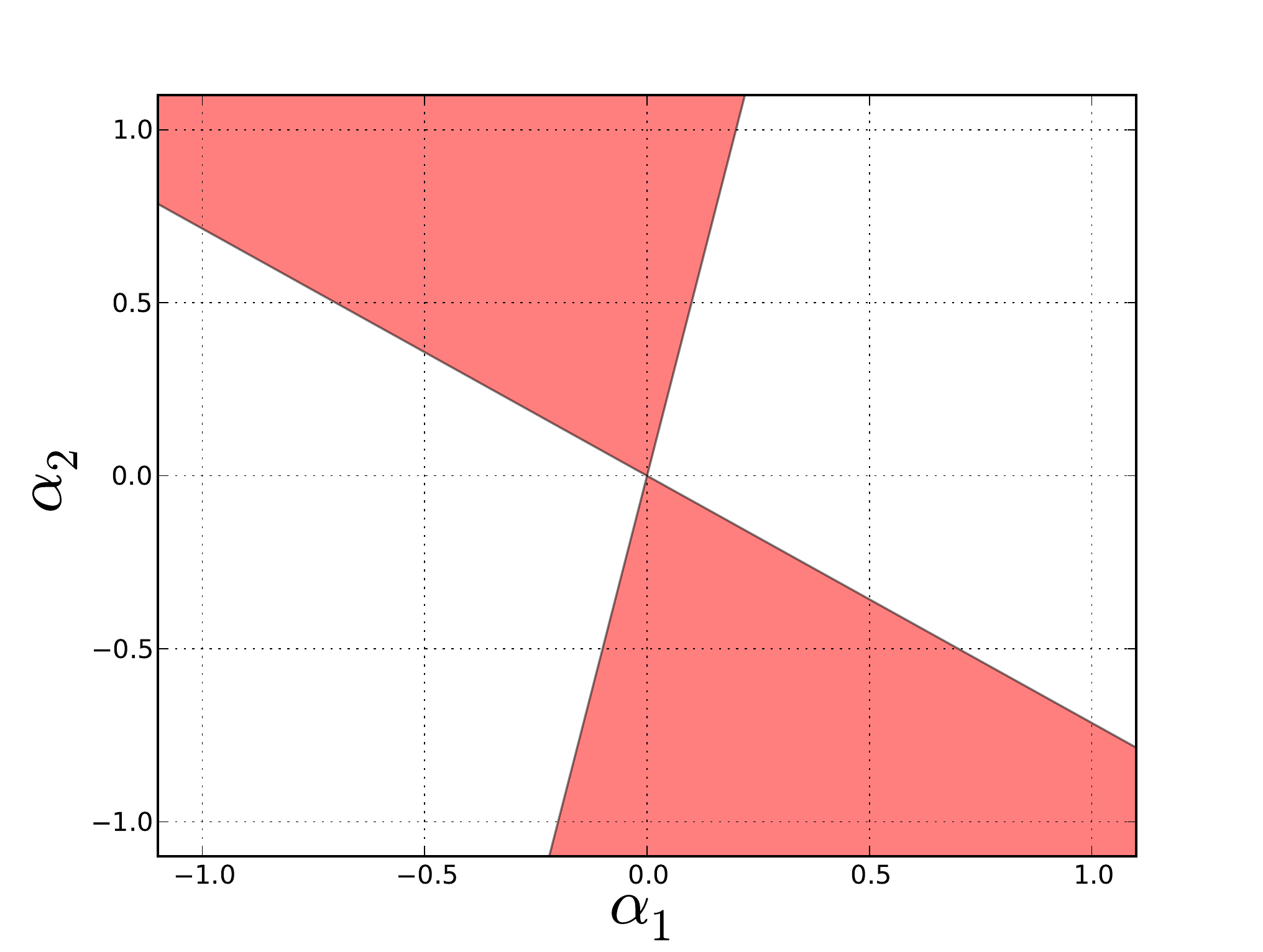}
\end{center}
\caption[]{The non-covered region of parameter space $(\alpha_1,\alpha_2)$ of a two dimensional model with $\sin\theta_{s_1,\T}=0$ and $\sin\theta_{s_2,\T}=0.8$}
\label{a1_a2}
\end{figure}
The opening angle of this wedge, $\beta$, is easily determined in terms of $\theta_2\equiv \theta_{s_2,\T}$ and is given by
\bea
\beta=-\tan^{-1}\frac{1}{\sin\theta_2+\cos\theta_2}-\tan^{-1}\frac{1}{\sin\theta_2-\cos\theta_2}\,, \nonumber
\eea
for $\theta_2\leq \pi/4$, and
\bea
\beta=\pi-\tan^{-1}\frac{1}{\sin\theta_2+\cos\theta_2}-\tan^{-1}\frac{1}{\sin\theta_2-\cos\theta_2}\,,\nonumber
\eea
for $\theta_2>\pi/4$ (note that $\theta_2$ can always be defined to be in the range $(0,\pi/2)$). The ratio $\beta/\pi$, which is an estimate of the non-covered parameter space, is plotted as a function of $\cos\theta_2$ in Fig. \ref{beta}. It is seen that for $\cos \theta_{s_2,\T} \leq 0.95$ more than about 10\% of the parameter space is not covered. One can therefore enlarge $\T$ by adding $t_2 = {s_2}_\perp/|{s_2}_\perp|$. This is how the orthogonal shape was originally constructed using $I_1=\dot\pi\pi_{,i}^2$, $I_2=\dot\pi^3$, and $\T=\{{\rm Equil.}\}$, where $\cos(s_1,\T)\simeq 1$ and $\cos(s_2,\T)\simeq 0.95$~\cite{Senatore:2009gt}.

\begin{figure}[h!]
\begin{center}
\includegraphics[width=10 cm, height= 8cm]{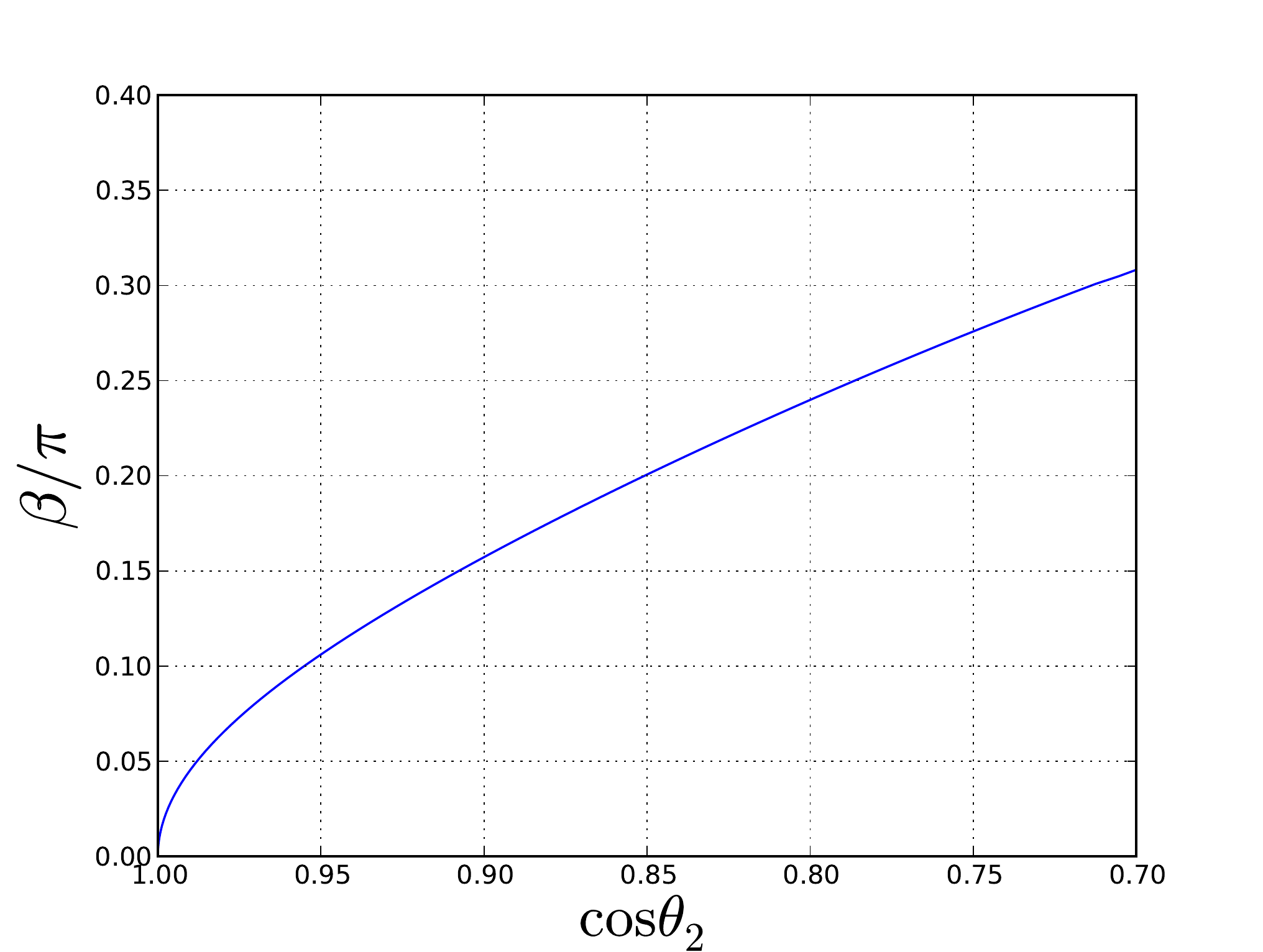}
\end{center}
\caption[]{An estimate of the non-covered fraction of parameter space of two-dimensional models as a function of $\cos \theta_2$.}
\label{beta}
\end{figure}


Before proceeding it is worth remarking that the choice of basis for the shapes has a dramatic effect on this sort of analysis. Even if $\cos(s_2,\T)>0.95$, in which case the above procedure does not enlarge $\T$, as long as $\cos(s_2,\T)\neq 1$, $s_1$ and $s_2$ span a two-dimensional space. Therefore, if we had chosen a different basis $(s_1,s_2')$ with $s_2'={s_2}_\perp/|{s_2}_\perp|$, the outcome would have been to enlarge $\T$. As mentioned in the text, we work with the normalized shapes of interactions that arise directly from the unitary gauge operators or linear combinations of them with order one coefficients. Choosing the basis with $s_2'={s_2}_\perp/|{s_2}_\perp|$ in the case  $\cos(s_2,\T)$ is very close to one would correspond to having large, not order one, coefficients in the unitary gauge Lagrangian. 

Let us now consider what would happen if $\SS$ contained an additional linearly dependent shape $s_3\simeq s_1$. The parameter space is now three dimensional, but the condition for having non-covered linear combination is the same as \eqref{coverage} except that $\alpha_1$ is now replaced with $\bar\alpha\equiv\alpha_1+\alpha_3$:
\bea
\label{coverage_bar}
|\alpha_2 {s_2}_\perp|>|\alpha_1+\alpha_3+\alpha_2 {s_2}_\parallel |\ .
\eea
There is no dependence on the perpendicular direction in $(\alpha_1,\alpha_3)$ plane; therefore, we get, as non-covered part, the same wedge cut out of the $3d$ parameter space except that the wedge is now squeezed along $\bar\alpha$ by a factor of $|\nabla_\alpha \bar\alpha|=[(\partial_{\alpha_1}\bar\alpha)^2+(\partial_{\alpha_3}\bar\alpha)^2]^{1/2}=\sqrt{2}$. To see this more clearly, choose a new orthonormal basis in $(\alpha_1,\alpha_3)$ plane given by
\bea
\hat \alpha_\pm=\frac{1}{\sqrt{2}}(\hat\alpha_1\pm\hat\alpha_2).
\eea
In this basis, we have $\bar \alpha = \sqrt{2}\alpha_+$, so \eqref{coverage_bar} becomes
\bea
\label{coverage+}
|\alpha_2 {s_2}_\perp|>|\sqrt{2}\alpha_+ + \alpha_2 {s_2}_\parallel |\,.
\eea
Since there is no dependence on $\alpha_-$, the problem of determining the non-covered fraction of parameter space has now manifestly reduced to the previous 2-dimensional case. However, for fixed $\alpha_2$, the condition \eqref{coverage+} is now satisfied in a range of size $\Delta \alpha_+= \sqrt{2} {s_2}_\perp \alpha_2$, which is smaller than \eqref{delta_alpha1} by a factor of $\sqrt{2}$. Generalizing, we therefore conclude that, for fixed $\T$, the non-covered portion of parameter space falls roughly as $1/\sqrt{\dim\SS_\T}$ for large $\SS_\T$. Here, $\SS_\T$ is defined as the subset of $\SS$ that lies inside $\T$, and $\dim \SS_\T$ is the number of shapes in our model that can be written as a linear combination of $\{t_i\}$. Hence, keeping all the other conditions fixed, models with fewer number of interactions are more likely to have larger non-covered parameter space. 

Finally consider the situation where we have added a second shape (say the Orthogonal shape) to $\T$ and there is a model with a new shape $s_3$: $\SS=\{s_1,s_2,s_3\}$, where $s_1$ and $s_2$ are generic linearly independent shapes fully inside $\T$. Similar to \eqref{coverage}, the condition that a linear combination $\sum \alpha_i s_i$ is not covered by $\T$ is given by
\bea
\label{coverage1}
|\alpha_3 {s_3}_\perp|>|\alpha_1s_1+\alpha_2 s_2+\alpha_3 {s_3}_\parallel |\ .
\eea
From the one dimensional experience, we expect that at fixed $\alpha_3$, $\alpha_{1,2}$ vary in a range $\Delta \alpha_{1,2}\sim {s_3}_\perp \alpha_3$ (c.f. \eqref{delta_alpha1}). However, since now $s_1$ and $s_2$ span a two dimensional plane, the non-covered portion of parameter space is roughly proportional to $\Delta \alpha_1 \Delta\alpha_2\propto {s_3}_\perp^2$, which is the square of the case with $\dim \T =1$. 

Although this scaling is not precise when $s_1$ and $s_2$ are not perpendicular, it can still be considered as an estimate of the relation between the non-covered parameter space and the dimensionality of $\T$. Generalizing this result to cases with larger $\dim \T$ and using ${s}_\perp=\sin \theta_{s,\T}$, we see that at fixed $\theta_{s,\T}$ the non-covered portion of the parameter space falls as 
\bea\label{eq:costhreshold}
\propto (\sin \theta_{s,\T})^{\dim \T}\,.
\eea
Hence, as $\dim \T$ is increased, the correlation $\cos \theta_{s_3,\T}$ must rapidly decrease to satisfy the criterion for adding a new template (that is, 10\% of the parameter space not being covered by $\T$). For $\dim\T =2$, this is about $\cos \theta_{\rm threshold}\simeq 0.8$. 

In our analysis, it is also useful to define a notion of effective template subspace $\T_{\rm eff}$. For each inflationary model, $\T_{\rm eff}$ is the subspace of $\T$ that is spanned by the shapes in that particular model. The above arguments are completely insensitive to the directions in $\T$ that are perpendicular to all of the shapes in a given model, so in the presence of such directions, all $\T$'s above must be replaced by $\T_{\rm eff}$. 

We, therefore, see that given a new shape $s$ with $\cos(s,\T)$ different from $1$, it is most likely that we need to expand $\T$ if there exist a model that contains $s$ and has $\dim\T_{\rm eff}=1$. Moreover, in this case we need $\cos(s,\T)<0.95$, so if we enlarge $\T$ to the extent that all shapes are more than $0.95$ correlated with $\T$, it is extremely likely that we cover more than 90\% of the parameter space of all models that are made from linear combination of those shapes. As we discussed, the requirement $\cos(s,\T)<0.95$ is obtained for $\dim\T_{\rm eff}=1$ and is therefore very conservative. For fixed maximum cosine the covered fraction of the parameter space gets larger as the dimensionality of $\T_{\rm eff}$ and ${\cal S}_\T$ grow. We next use the above results to analyze the space of technically natural models as the number of derivatives is increased. 

\section{\label{templates} Order by order search for new templates}

Our practical strategy to decide if we need to include a new template can be summarized as follows:

a) Since we start the analysis from the 2-dimensional $\T=\{{\rm Equil.,Ortho.}\}$, and because at higher derivative levels technically natural models usually have to contain many operators that are connected to one another by loops, we expect the effective dimension of template space $\T_{\rm eff}$ to be naturally larger than one. Even if it is 1, there is still the suppression $1/\sqrt{\dim \SS_{\T}}$ that we need to consider, as we discussed above. So instead of $\cos(s,\T)<0.95$, we adopt the more stringent condition of $\cos(s,\T)<0.9$, and consider only these shapes $s$ as {\it candidate} new shapes. Whenever an inflationary model contains such a candidate shape in addition to some other shapes (or equivalently interactions), there is a probability that the overall shape of the model be less than $0.7$ correlated with $\T$ for some range of parameters.

b) We enlarge $\T$ only if we can find at least one technically natural model such that more than about 10\% of its parameter space is not covered by $\T$. Indeed, depending on the dimensionality ${\cal S}$ of the model and its $\T_{\rm eff}$, the criterion of $\cos(s,\T)<0.9$ might not be enough to ensure a 10\%-fraction of non-covered parameter space. In practice, however, we are unable to check explicitly all the models due to their large number. Therefore there is a risk of missing to include a new template simply because we are unable to identify one such a model. However, the following point make us reasonably confident that this is not the case.

c) At the end of our investigation up to the 9-derivative level, we find a 4-dimensional~$\T$. Only 3 out of 140 individual shapes that can appear in technically natural models have a correlation less than $0.95$ with $\T$. As we are going to explain next in some detail, even for models involving these three shapes, the arguments of appendix \ref{geo} suggest that we are likely to cover more than 90\% of the parameter space of technically natural models that are considered here.

We start at 6-derivative level since it was shown in section \ref{indi} that no new shape can appear at 4- or 5-derivative level in models that satisfy the criteria of appendix \ref{natural}. Following step (a), we need to find all shapes that appear at 6-derivative level and have cosine smaller than $0.9$ with $\T=\{{\rm Equil.,Ortho.}\}$. There is only one such shape at this level, and its individual correlations are shown in Table \ref{6der}. The correlation is low enough that this shapes become a candidate shape. 

\begin{table}[t]
\begin{center}
\begin{tabular}{| c | c | c | c| }
\hline			
cosine with & Equil & Ortho  & $\T$  \\ \hline
$\pi_{,ij}\pi_{,jk}\pi_{,ik}$ &0.07 & - 0.81 & 0.84 \\\hline
\end{tabular}
 \end{center}
\caption{Candidate new shapes at 6-derivative}
\label{6der}
\end{table}

Following step (b), we need to find a model that contains this operator and more than 10\% of its parameter space is not covered by $\T$. It seems, however, that no such model exists. First notice that in a technically natural model the interaction $\pi_{,ij}\pi_{,jk}\pi_{,ik}$ must come together with $\d^2\pi\pi_{,ij}^2$ and $(\d^2\pi)^3$, so the minimal model with our candidate new shape is
\bea
\label{6derL}
\mathcal{L}^{(3)}=\frac{1}{\Lambda^5}(a_1\pi_{,ij}\pi_{,jk}\pi_{,ik}+a_2\d^2\pi\pi_{,ij}^2+a_3(\d^2\pi)^3)\ .
\eea
By varying the coefficients of this model in an order one range (see footnote \ref{(-1,1)}), it is easy to see that the fraction of parameter space of this model that is not covered by $\T=\{ {\rm Equil., Ortho.}\}$ is negligible. When we add more shapes to the model, this fraction grows in some cases~\footnote{This fact seems to contradict the arguments we gave in Appendix~\ref{geo}. However, those augments apply when  the vectors are not particularly aligned, and so they are expected to be true in the limit in which the dimensionality of the space is large.}.  For instance, in the following model
\bea
\mathcal{L}^{(3)}=\frac{1}{\Lambda^5}(a_1\ddot\pi^3+a_2 \ddot\pi\pi_{,ij}^2 +a_3\d^2\pi\pi_{,ij}^2 +a_4(\d^2\pi)^3+a_5\pi_{,ij}\pi_{,jk}\pi_{,ik}+a_6\d^2\pi \ddot\pi^2)\ ,
\eea
about 5\% of the parameter space is not covered by $\T$. Yet we could not find any model with a larger fraction of non-covered parameter space, and it seems unlikely that any such model exists since the dimensionality of $\SS$ is already grown larger than 3 (as explained above \eqref{6derL}); therefore, it is very probable that $\dim \T_{\rm eff}>1$ or the $1/\sqrt{\dim\SS_\T}$ suppression becomes important, and the above 5\% seems to be the largest fraction that could be obtained. We, therefore, move to the next level without extending $\T$. Nonetheless, this 5\% of parameter space will eventually be covered by our final 4-dimensional $\T$ since the correlation of $\pi_{,ij}\pi_{,jk}\pi_{,ik}$ with that $\T$ is larger than $0.95$.

\begin{table}[t]
\begin{center}
\begin{tabular}{| c | c | c |  c |}
\hline			
cosine with & Equil & Ortho  & $\T$  \\ \hline
$\tdot\pi\dot\pi_{,i}^2$ & 0.47 & 0.75 & 0.82 \\\hline
$\tdot\pi\pi_{,ij}^2$ & 0.26 & -0.65  & 0.76\\\hline
$\pi_{,ij}\ddot\pi_{,i}\dot\pi_{,j}$ & 0.65 & -0.45 & 0.88 \\\hline
$\tdot\pi(\d^2\pi)^2$ & 0.66 & 0.70 & 0.87 \\\hline
$\pi_{,ijk}\pi_{,ij}\dot\pi_{,k}$ & -0.72 & 0.36 & 0.89 \\\hline
$\d^2\dot\pi\pi_{,ij}^2$ & 0.72 & -0.36 & 0.89 \\\hline
$\dot\pi_{,ij}\pi_{,jk}\pi_{,ik}$ & 0.072 & -0.81 & 0.84 \\\hline
\end{tabular}
 \end{center}
\caption{Candidate new shapes at 7-derivative level}
\label{7der}
\end{table}

At 7-derivative level, in addition to those listed in Table \ref{6der}, there are many candidate new shapes shown in Table \ref{7der}. We find that about 20\% of the parameter space of the technically natural model
\bea
\label{7derL}
\mathcal{L}^{(3)}=\frac{1}{\Lambda^5}(a_1 \tdot\pi\pi_{,ij}^2+a_2 H \pi_{,ij}\pi_{,jk}\pi_{,ik}+a_3 H\d^2\pi\pi_{,ij}^2+a_4 H(\d^2\pi)^3)\ ,
\eea
is not covered by $\T=\{ {\rm Equil.,Ortho.}\}$. Therefore, one can confidently add a template, that we arbitrarily choose to be based on the shape of $\tdot\pi\pi_{,ij}^2$ to $\T$. After this, the correlation of all technically permitted shapes at 7-derivative level with the new three-dimensional $\T$ grows above $0.9$. So we can move to the next level. 

\comment{with the exception of $\d^2\pi\ddot\pi_{,i}\dot\pi_{,i}$ that has a cosine of $0.93$. This operator, however, must come with $(\d^2\pi)^3$ and $\d^2\pi\dot\pi_{,i}^2$ and one can check that the parameter space of this model is almost completely covered by $\T$. Moreover, models that contain these three operators plus a few other 7 or lower derivative operators are unlikely to have $\dim\T_{\rm eff}=1$, so according to the argument of appendix \ref{geo} they are also mostly covered by $\T$.}

With the new 3-dimensional $\T$, there is no candidate new shape (with $\cos(s,\T)<0.9$) that can appear in a technically natural model with leading 8-derivative interactions. However, at 9-derivative level we have several candidates listed in table \ref{9der}. 

\begin{table}[t]
\begin{center}
\begin{tabular}{| c | c | c | c|  c |} \hline			
cosine with & Equil & Ortho & $\tdot\pi\pi_{,ij}^2$ & $\T$  \\ \hline
$\tdot\pi\ddot\pi_{,i}^2$ & -0.24 & 0.78 &  0.68 & 0.88 \\ \hline
$\tdot\pi\dot\pi_{,ij}^2$ & 0.65 & 0.30 & 0.34 & 0.89 \\\hline
$\tdot\pi\d^2\pi_{,i}\ddot\pi_{,i}$ & -0.74 & 0.19 & 0.26 & 0.87 \\\hline
$\tdot\pi\pi_{,ijk}^2$& 0.45 & 0.37 & 0.53 & 0.87 \\\hline
$\ddot\pi_{,i}\pi_{,ijk}\pi_{,jk}$  & 0.15 & 0.70 & 0.19 & 0.81 \\\hline
$\dot\pi_{,ij}\d^2\pi_{,i}\ddot\pi_{,j}$ & 0.29 & 0.83 & 0.29 & 0.90 \\\hline
$\dot\pi_{,ij}\pi_{,ikl}\pi_{,jkl}$  & -0.05 & -0.76 & -0.25 & 0.89 \\\hline
\end{tabular}
 \end{center}
\caption{Candidate new shapes at 9-derivative level}
\label{9der}
\end{table}

Again, there is some freedom as how to enlarge $\T$ since adding each of the new shapes causes the correlation of all the others to increase. Note, also, that at this level technically natural models often contain many shapes, which are connected by loops. This makes the search for models with large non-covered parameter space rather difficult. An example of a model with a relatively large non-covered parameter space is 
\bea
\mathcal{L}^{(3)}=\frac{1}{\Lambda^8}(a_1\tdot\pi\ddot\pi_{,i}^2+a_2 \tdot\pi^3+a_3 H^3 \pi_{,ij}\pi_{,jk}\pi_{,ik}).
\eea
About 8.4\% of the parameter space of this model is not covered by the three dimensional basis. Although this is slightly below the 10\% criterion, we still enlarge $\T$ by adding a new template based on $\tdot\pi\ddot\pi_{,i}^2$, also because the number of shapes with correlation less than $0.95$ with $\T=\{ {\rm Equil.,Ortho.,}\tdot\pi\pi_{,ij}^2\}$ is still statistically significant (17/138). Another ancillary reason to justify this extension is that we are not continuing the analysis to higher orders, and so adding a new template might accidentally cover for potential ones that might appear at higher orders.

\begin{table}[t]
\begin{center}
\begin{tabular}{| c | c | c | c|  c |c|}
\hline			
cosine with & Equil & Ortho &$\tdot\pi\pi_{,ij}^2$&$\tdot\pi\ddot\pi_{,i}^2$& $\T$  \\ \hline
$\ddot\pi_{,i}\pi_{,ijk}\pi_{,jk}$ & 0.15 & 0.70 & 0.19 & 0.59 & 0.82 \\ \hline
$\ddot\pi_{,i}\pi_{,ijk}\dot\pi_{,jk}$ & -0.40 & 0.34 & -0.04 & 0.48 & 0.92 \\ \hline
$\dot\pi_{,ij}\pi_{,ikl}\pi_{,jkl}$ & -0.05 & -0.76 & -0.25 & -0.49 &0.94 \\ \hline
\end{tabular}
 \end{center}
\caption{Remaining candidates at 9-derivative level}
\label{rem9der}
\end{table}

We think this is a reasonable point to stop extending $\T$ since except for three shapes that are shown in Table \ref{rem9der}, all other shapes have a correlation larger than $0.95$ with the 4-dimensional $\T$. Thus, the arguments of appendix \ref{geo} based on the dimensionality of the candidate natural model  suggest that $\T$ covers more than 90\% of the parameter space of all the models, even those which include the three shapes of table \ref{rem9der}. Take, for instance, $I_8\equiv\ddot\pi_{,i}\pi_{,ijk}\pi_{,jk}$ which has the smallest correlation with $\T$. It must come with an $H$ suppression compared to some 9-derivative operator, say $I_9=\tdot\pi^3$, otherwise it generates the cubic $\pi_{,ij}^3$ interaction without suppression. Then we must also include the 8-derivative interaction $H \tdot\pi^2\d^2\pi$ which will be generated by a loop with two $I_9$ and one $I_8$. Moreover, loops with three $I_8$ generate $H^3\d^2\pi\pi_{,ij}^2,H^3(\d^2\pi)^3$, and $H^3\pi_{,ij}\pi_{,jk}\pi_{,ik}$. Therefore, $\dim T_{\rm eff}$ is very unlikely to be small, and there is a large $1/\sqrt{\dim \SS_T}$ suppression~\footnote{For example, it is enough for the model to have $\dim \T_{\rm eff}$ to be greater or equal to 2, for considering, according to~(\ref{eq:costhreshold}),  this shape to be covered.}. Thus, we expect more than 90\% of the parameter space of any model which includes $I_8$ to be covered by $\T=\{{\rm Equil.,Ortho.,}\tdot\pi\pi_{,ij}^2,\tdot\pi\ddot\pi_{,i}^2\}$. Similar arguments apply to the other two operators in Table \ref{rem9der}. Explicit computations of the non-covered parameter space in a handful of models were in agreement with the above expectations.

Let us conclude with a general remark. In this analysis we made a lot of rather subjective choices, and we hope that the inevitability of them, once one realizes the technical naturalness of models with higher derivative interactions, is appreciated by the reader. Making different choices would lead to different outcomes for templates, and this may be interpreted as a sign of unreliability of the final result. However, we should emphasize that there is a huge degree of degeneracy in this problem. Firstly, all derivative interactions reduce to those of the form (\ref{t},\ref{i}), which is a reduction from about 140 operators at 9-derivative level to only 9. Secondly, even that 9-dimensional basis is redundant in practice, and we argue a 4-dimensional $\T$ is sufficient to cover the space of these models. This 4-dimensional basis is not unique, as one can obviously redefine the basis, and most importantly  we can imagine a different set of criteria may even make it 5-dimensional; but, regardless of this rather small arbitrariness, this set of templates seems to be remarkably reliable in exploring the signatures of the possible inflationary models with the observational data.



\begin{thebibliography}{99}

 \bibitem{Cheung:2007st}
  C.~Cheung, P.~Creminelli, A.~L.~Fitzpatrick, J.~Kaplan, L.~Senatore,
  ``The Effective Field Theory of Inflation,''
  JHEP {\bf 0803}, 014 (2008).
  [arXiv:0709.0293 [hep-th]].
  



\bibitem{Creminelli:2005hu}
  P.~Creminelli, A.~Nicolis, L.~Senatore, M.~Tegmark and M.~Zaldarriaga,
  ``Limits on non-gaussianities from wmap data,''
  JCAP {\bf 0605} (2006) 004
  [astro-ph/0509029].


\bibitem{Senatore:2009gt}
  L.~Senatore, K.~M.~Smith and M.~Zaldarriaga,
   ``Non-Gaussianities in Single Field Inflation and their Optimal Limits from
  the WMAP 5-year Data,''
  JCAP {\bf 1001}, 028 (2010)
  [arXiv:0905.3746 [astro-ph.CO]].

\bibitem{Alishahiha:2004eh}
  M.~Alishahiha, E.~Silverstein and D.~Tong,
  ``DBI in the sky,''
  Phys.\ Rev.\ D {\bf 70} (2004) 123505
  [hep-th/0404084].





\bibitem{Komatsu:2003iq}
  E.~Komatsu, D.~N.~Spergel and B.~D.~Wandelt,
  ``Measuring primordial non-Gaussianity in the cosmic microwave background,''
  Astrophys.\ J.\  {\bf 634} (2005) 14
  [astro-ph/0305189].

  
\bibitem{Lyth:2002my}
  D.~H.~Lyth, C.~Ungarelli and D.~Wands,
  ``The Primordial density perturbation in the curvaton scenario,''
  Phys.\ Rev.\ D {\bf 67} (2003) 023503
  [astro-ph/0208055].


\bibitem{Zaldarriaga:2003my}
  M.~Zaldarriaga,
  ``Non-Gaussianities in models with a varying inflaton decay rate,''
  Phys.\ Rev.\ D {\bf 69} (2004) 043508
  [astro-ph/0306006].


\bibitem{Senatore:2010wk}
  L.~Senatore and M.~Zaldarriaga,
  ``The Effective Field Theory of Multifield Inflation,''
  JHEP {\bf 1204} (2012) 024
  [arXiv:1009.2093 [hep-th]].


\bibitem{Fergusson:2009nv}
  J.~R.~Fergusson, M.~Liguori and E.~P.~S.~Shellard,
  ``General CMB and Primordial Bispectrum Estimation I: Mode Expansion, Map-Making and Measures of $f_{\rm NL}$,''
  Phys.\ Rev.\ D {\bf 82} (2010) 023502
  [arXiv:0912.5516 [astro-ph.CO]].

  
  
\bibitem{Ade:2013ydc}
  P.~A.~R.~Ade {\it et al.}  [Planck Collaboration],
  ``Planck 2013 Results. XXIV. Constraints on primordial non-Gaussianity,''
  arXiv:1303.5084 [astro-ph.CO].


\bibitem{Carrasco:2012cv}
  J.~J.~M.~Carrasco, M.~P.~Hertzberg and L.~Senatore,
  ``The Effective Field Theory of Cosmological Large Scale Structures,''
  JHEP {\bf 1209} (2012) 082
  [arXiv:1206.2926 [astro-ph.CO]].

  J.~J.~M.~Carrasco, S.~Foreman, D.~Green and L.~Senatore,
  ``The Effective Field Theory of Large Scale Structures at Two Loops,''
  arXiv:1310.0464 [astro-ph.CO].

  R.~A.~Porto, L.~Senatore and M.~Zaldarriaga,
  ``The Lagrangian-space Effective Field Theory of Large Scale Structures,''
  JCAP {\bf 1405} (2014) 022
  [arXiv:1311.2168 [astro-ph.CO]].

  L.~Senatore and M.~Zaldarriaga,
  ``The IR-resummed Effective Field Theory of Large Scale Structures,''
  arXiv:1404.5954 [astro-ph.CO].

  R.~E.~Angulo, S.~Foreman, M.~Schmittfull and L.~Senatore,
  ``The One-Loop Matter Bispectrum in the Effective Field Theory of Large Scale Structures,''
  arXiv:1406.4143 [astro-ph.CO].

  T.~Baldauf, L.~Mercolli, M.~Mirbabayi and E.~Pajer,
  ``The Bispectrum in the Effective Field Theory of Large Scale Structure,''
  arXiv:1406.4135 [astro-ph.CO].


\bibitem{LopezNacir:2011kk}
  D.~Lopez Nacir, R.~A.~Porto, L.~Senatore and M.~Zaldarriaga,
  ``Dissipative effects in the Effective Field Theory of Inflation,''
  JHEP {\bf 1201} (2012) 075
  [arXiv:1109.4192 [hep-th]].

\bibitem{Creminelli:2006xe}
  P.~Creminelli, M.~A.~Luty, A.~Nicolis and L.~Senatore,
  ``Starting the Universe: Stable Violation of the Null Energy Condition and Non-standard Cosmologies,''
  JHEP {\bf 0612} (2006) 080
  [hep-th/0606090].

\bibitem{Gubitosi:2012hu}
  G.~Gubitosi, F.~Piazza and F.~Vernizzi,
  ``The Effective Field Theory of Dark Energy,''
  JCAP {\bf 1302} (2013) 032
   [JCAP {\bf 1302} (2013) 032]
  [arXiv:1210.0201 [hep-th]].

\bibitem{Bartolo:2010bj}
  N.~Bartolo, M.~Fasiello, S.~Matarrese and A.~Riotto,
  ``Large non-Gaussianities in the Effective Field Theory Approach to Single-Field Inflation: the Bispectrum,''
  JCAP {\bf 1008} (2010) 008
  [arXiv:1004.0893 [astro-ph.CO]].





\bibitem{Creminelli:2010qf}
  P.~Creminelli, G.~D'Amico, M.~Musso, J.~Norena and E.~Trincherini,
  ``Galilean symmetry in the effective theory of inflation: new shapes of
  non-Gaussianity,''
  JCAP {\bf 1102}, 006 (2011)
  [arXiv:1011.3004 [hep-th]].
  
  

  
\bibitem{Flauger:2010ja}
  R.~Flauger and E.~Pajer,
  ``Resonant Non-Gaussianity,''
  JCAP {\bf 1101} (2011) 017
  [arXiv:1002.0833 [hep-th]].


\bibitem{Nicolis:2008in}
  A.~Nicolis, R.~Rattazzi and E.~Trincherini,
  ``The Galileon as a local modification of gravity,''
  Phys.\ Rev.\ D {\bf 79} (2009) 064036
  [arXiv:0811.2197 [hep-th]].

  
\bibitem{Adams:2006sv}
  A.~Adams, N.~Arkani-Hamed, S.~Dubovsky, A.~Nicolis and R.~Rattazzi,
  ``Causality, analyticity and an IR obstruction to UV completion,''
  JHEP {\bf 0610} (2006) 014
  [hep-th/0602178].

  
  
\bibitem{Babich:2004gb}
  D.~Babich, P.~Creminelli and M.~Zaldarriaga,
  ``The Shape of non-Gaussianities,''
  JCAP {\bf 0408} (2004) 009
  [astro-ph/0405356].

\bibitem{Maldacena:2002vr}
  J.~M.~Maldacena,
  ``Non-Gaussian features of primordial fluctuations in single field
  inflationary models,''
  JHEP {\bf 0305} (2003) 013
  [arXiv:astro-ph/0210603].



  
 
\bibitem{resonant} 
  S.~R.~Behbahani, A.~Dymarsky, M.~Mirbabayi and L.~Senatore,
  ``(Small) Resonant non-Gaussianities: Signatures of a Discrete Shift Symmetry in the Effective Field Theory of Inflation,''
  arXiv:1111.3373 [hep-th].

\bibitem{Daniel_Siavosh} 
  S.~R.~Behbahani and D.~Green,
  ``Collective Symmetry Breaking and Resonant Non-Gaussianity,''
  arXiv:1207.2779 [hep-th].

\bibitem{4point}
 L.~Senatore, M.~Zaldarriaga,
  ``A Naturally Large Four Point Function in Single Field Inflation,''
  JCAP {\bf 1101}, 003 (2011).
  [arXiv:1004.1201 [hep-th]].


\bibitem{Smith:2006ud}
  K.~M.~Smith and M.~Zaldarriaga,
  ``Algorithms for bispectra: Forecasting, optimal analysis, and simulation,''
  Mon.\ Not.\ Roy.\ Astron.\ Soc.\  {\bf 417} (2011) 2
  [astro-ph/0612571].

\bibitem{Wang:1999vf}
  L.~-M.~Wang and M.~Kamionkowski,
  ``The Cosmic microwave background bispectrum and inflation,''
  Phys.\ Rev.\ D {\bf 61} (2000) 063504
  [astro-ph/9907431].



\bibitem{Creminelli:2006rz}
  P.~Creminelli, L.~Senatore, M.~Zaldarriaga and M.~Tegmark,
  ``Limits on $f_{\rm NL}$ parameters from WMAP 3yr data,''
  JCAP {\bf 0703} (2007) 005
  [astro-ph/0610600].


\bibitem{Yadav:2007ny}
  A.~P.~S.~Yadav, E.~Komatsu, B.~D.~Wandelt, M.~Liguori, F.~K.~Hansen and S.~Matarrese,
  ``Fast Estimator of Primordial Non-Gaussianity from Temperature and Polarization Anisotropies in the Cosmic Microwave Background II: Partial Sky Coverage and Inhomogeneous Noise,''
  Astrophys.\ J.\  {\bf 678} (2008) 578
  [arXiv:0711.4933 [astro-ph]].

\bibitem{Lewis:1999bs}
  A.~Lewis, A.~Challinor and A.~Lasenby,
  ``Efficient computation of CMB anisotropies in closed FRW models,''
  Astrophys.\ J.\  {\bf 538} (2000) 473
  [astro-ph/9911177].

\bibitem{Bennett:2012zja}
  C.~L.~Bennett {\it et al.}  [WMAP Collaboration],
  ``Nine-Year Wilkinson Microwave Anisotropy Probe (WMAP) Observations: Final Maps and Results,''
  Astrophys.\ J.\ Suppl.\  {\bf 208} (2013) 20
  [arXiv:1212.5225 [astro-ph.CO]].


\end{thebibliography}
\end{document}